\documentclass[11pt]{article}

\usepackage[top=0.8in, bottom=0.78in, left=0.87in, right=0.87in]{geometry}
\usepackage[table]{xcolor}
\usepackage{setspace}
\usepackage[T1]{fontenc}
\usepackage{times}
\usepackage{booktabs}
\usepackage{rotating}
\usepackage{graphicx}
\usepackage[section]{placeins} 
\usepackage[large, bf]{caption}
\usepackage{subcaption}
\usepackage{pdfpages}
\usepackage{palatino}
\usepackage{pdflscape}
\usepackage{textcomp}
\usepackage{longtable}
\usepackage{nicefrac}
\usepackage{adjustbox}	
\usepackage[hyphens]{url}

\usepackage{natbib}
\bibpunct{(}{)}{;}{a}{,}{,}

\usepackage{amsmath, amsfonts, amssymb, amsthm}

\usepackage{mathpazo} 
\usepackage[pdftex]{hyperref}
\hypersetup{colorlinks, citecolor=blue, filecolor=blue, linkcolor=blue, urlcolor=blue}

\def\urltilda{\kern -.15em\lower .7ex\hbox{\~{}}\kern .04em}

\usepackage{titlesec}

\titlespacing*{\section}{0pt}{1.5ex plus 1ex minus .2ex}{0.8ex plus .2ex}
\titlespacing*{\subsection}{0pt}{1.2ex plus 1ex minus .2ex}{0.8ex plus .2ex}

\usepackage{dcolumn}
\newcolumntype{d}[0]{D{.}{.}{5}}
\usepackage{mathtools}
\usepackage{tikz}
\usepackage{graphicx}
\usepackage[space]{grffile}

\newtheorem*{rem*}{\protect\remarkname}
\providecommand{\remarkname}{Remark}

\usepackage[official]{eurosym}
\usetikzlibrary{arrows,positioning} 
\usetikzlibrary{decorations.pathreplacing}
\usetikzlibrary{decorations.pathmorphing}
\usetikzlibrary{intersections}
\usepackage{float}
\usepackage[american]{babel}

\usepackage{tikz}
\usetikzlibrary{patterns}
\usepackage{adjustbox}

\usepackage{makecell}

\usepackage{algorithm}
\usepackage{algcompatible}

\algnewcommand\algorithmicreturn{\textbf{return}}
\algnewcommand\RETURN{\State \algorithmicreturn}%

  {
      \theoremstyle{plain}
      
  }
\usepackage{booktabs}
\usepackage{graphicx}
\usepackage{caption}
\usepackage{subcaption}
\usepackage{cleveref}
\usepackage{stfloats}
\usepackage{tabulary}
\usepackage{multirow}
\usepackage{algpseudocode}
\usepackage{algorithm}
\newcommand{\indep}{\perp \!\!\! \perp}
\usepackage{blkarray}

\algnewcommand\algorithmicforeach{\textbf{for each}}
\algdef{S}[FOR]{ForEach}[1]{\algorithmicforeach\ #1\ \algorithmicdo}

\begin{document}


\title{\textbf{\LARGE{Personalized Recommendations in EdTech: Evidence from a Randomized Controlled Trial}}\thanks{ We are grateful to Deepak Agarwal and his team at Stones2Milestones for collaboration on this project. We would like to thank Rina Park and Shanjukta Nath for helpful comments and suggestions. Sarthak Kanodia and Jazon Jiao provided excellent research assistance. The Golub Capital Social Impact Lab at Stanford Graduate School of Business provided funding for this research.}\\ \vspace{0.2cm} \normalsize }

\author{Keshav Agrawal\thanks{Stanford Graduate School of Business, keshavag@stanford.edu} \and Susan Athey\thanks{Stanford Graduate School of Business, athey@stanford.edu}\and Ayush Kanodia\thanks{Stanford Computer Science, akanodia@stanford.edu}\and Emil Palikot\thanks{Stanford Graduate School of Business, palikot@stanford.edu.}}
\date{\today}

\maketitle

\thispagestyle{empty} 
\setcounter{page}{0}
\begin{abstract}
 		We study the impact of personalized content recommendations on the usage of an educational app for children. In a randomized controlled trial, we show that the introduction of personalized recommendations increases the consumption of content in the personalized section of the app by approximately 60\%. We further show that the overall app usage increases by 14\%, compared to the baseline system where human content editors select stories for all students at a given grade level. The magnitude of individual gains from personalized content increases with the amount of data available about a student and with preferences for niche content: heavy users with long histories of content interactions who prefer niche content benefit more than infrequent, newer users who like popular content. To facilitate the move to personalized recommendation systems from a simpler system, we describe how we make important design decisions, such as comparing alternative models using offline metrics and choosing the right target audience.
\end{abstract}
\newpage{}

\section{Introduction}

Recommendation systems, the algorithms that determine which pieces of content will be displayed to each user, have been widely deployed in online services. In particular, personalization of content recommendations has contributed to the success of some of the most valuable companies in the world; for example, \emph{Netflix} and \emph{Spotify} in the entertainment industry and \emph{Amazon} in online retail. These companies are at the forefront of developing these innovative algorithms and reap high benefits from implementing them.\footnote{For example, \cite{gomez2015netflix} assess the annual value of personalized content recommendations at \emph{Netflix} to be over \$1B. \cite{schrage2020recommendation} discusses the value of recommendation systems in \emph{Alibaba} and \emph{Amazon}.} The success and profitability of these algorithms raise the question of whether they could be valuable in other settings as well. Such a setting of particular promise and poential impact is education.

\par The benefits of personalization depend on the fundamental preferences of the users (e.g., students). Human curation or simple popularity-based algorithms may be sufficient if their preferences are homogeneous. In contrast, if users have diverse preferences, adjusting content selection to their tastes can be beneficial. The degree of preference heterogeneity might differ across sectors. Therefore, empirical evidence is required to understand the importance of personalization in an educational setting. However, in many educational settings (e.g., mobile educational apps), the user base may be orders of magnitude smaller than popular entertainment applications, and so it is unclear whether methods for creating personalized recommendations that have been successful in other sectors would be effective in an educational setting.

\par Personalization of content selection in the digital education context typically takes the form of either assigning learning materials at the difficulty level right for the specific user or adjusting the content's type to match the user's preferences. In this paper, we focus on the latter.\footnote{See \cite{escueta2020upgrading} for a review of the literature on the impact of personalization of learning content difficulty on learning outcomes.} It is not a priori clear that personalized content increases app usage. Notably, learners might engage with digital education products following a specific routine or, in the case of children, the requirements of parents or teachers. On the other hand, if personalization of content selection increases engagement with educational material, educational apps can become more effective learning tools. However, the publicly available evidence of the impact of personalized content recommendations in education is limited.

\par This paper aims to fill some of these gaps by providing evidence from a randomized controlled trial (RCT) designed to measure the impact of the introduction of personalized recommendations in place of recommendations curated manually by editors (an "editorial-based system") into \emph{Freadom}, an educational mobile and web application for Indian children learning to read in English. In particular, we conducted a randomized experiment over a two-week period in which we changed the way an app's content was selected. The control group was exposed to stories based on the status quo, editorial-based system. In contrast, the treatment group received personalized recommendations in a section of the app.

\par We find that personalization leads to a substantial increase in user engagement with the app compared to the editorial-based system: our estimate of the increase in usage of the personalized section is 60\% (S.E. 17\%). Since the personalized content was shown in one section of the app, it is possible that users might shift from consuming editorial-based content in other sections of the app to consuming personalized content in the personalized section without increasing overall engagement. To assess this possibility, we also estimate the effect of introducing personalization on app usage across all sections of the app, and we estimate an increase of approximately 14\% (S.E. 7\%).

\par  Although the ultimate goal of increasing engagement is to improve student outcomes, in this paper, we do not have measures of the impact on learning. In particular, we cannot measure the opportunity cost of students' time. For example, if the time spent engaging with this educational app substitutes for non-educational apps or other leisure, it might be more likely to have an overall positive impact. However, we have no direct evidence that this is the case.

\par The app \emph{Freadom} is developed by \emph{Stones2Milestones (S2M)}, and is targeted at children aged 3 to 12. The app aims to address challenges faced by students without access to appropriate English-language reading material at home, especially for students whose parents lack English-language skills. The app's main content consists of short, illustrated stories appropriate for the targeted age groups. Each story is a self-contained learning unit that generally consists of a reading part and a quiz. \emph{Freadom} is mostly used on smartphones, where the display page consists of various sections (more precisely, trays), each of which contains a sequence of stories sorted by an algorithm. Sections (trays) of the app are labeled with different names and display stories according to a specific algorithm. When we conducted this research, the recommendation algorithms were not personalized: stories were selected through manual curation or simple algorithms.

To design, implement, and evaluate a personalized recommendation system, we proceeded in the following order. First, we evaluated alternative versions of recommendation systems using historical data. Second, we implemented one of the alternatives in the context of a randomized experiment, in which we allocate a randomly selected subset of users to the personalized recommendation system. Third, we evaluated the benefits of the system compared to the status quo.

To select a recommendation system for deployment, we compared several alternative approaches to building recommendation systems, including systems based on a fixed effects model, a two-way fixed effects model (``popularity-based model"), and a collaborative filtering model (matrix factorization). We compared these systems and ultimately selected an approach based on matrix factorization \citep{mnih2007probabilistic, rendle2010factorization}. Comparing alternatives can be challenging. Even though historical data about user-story interactions can be used to assess alternative models, historical data does not capture the effects of sustained exposure to a new personalized recommendation system, since, in historical data, most individuals are typically exposed to only a few stories that a counterfactual recommendation system would have recommended. In addition, relative to policy evaluation in settings that have only a few treatment arms (for which doubly robust methods are recommended, e.g., \cite{kitagawa2018should, athey2021policy, zhou2022offline}), in the setting we study, there are approximately 2200 stories shown in a given week, so there may be limited overlap between a counterfactual policy and historical data. In other words, few users were historically assigned to the stories they would have been assigned if a counterfactual recommendation system had been in place. Furthermore, the user-story interactions that do overlap between the recommendation system in the historical data and the counterfactual system are not representative of the interactions that would occur in the counterfactual recommendation system.

The challenges of using historical data to evaluate a recommendation policy motivate the randomized experiment. We implemented the recommendation system based on matrix factorization and allocated a randomly selected subset of users to receive recommendations from this system. We deployed the personalized recommendation system in the section of the app called \emph{Recommended Story}. Users in the control group continued to experience the status quo (the editorial policy). Since matrix factorization requires substantial user history to perform well, the experiment only included users who interacted with at least sixty stories before the start of the experiment.

Using the experimental data, we compare outcomes between the treatment and control group in what we refer to as ``on-policy'' evaluation. \emph{Total Story Engagement}, defined as the sum of the \emph{Story Engagement}s from all user-story interactions during the experiment, is the primary outcome we investigate. \emph{Story Engagement} is a constructed metric that assigns a value of 1.0 if a user finished a story, 0.5 if a user started the story but did not finish it, and 0.3 if the user clicked on the story to view the description but did not start it. If a user does not click on the story, we assign the value of 0.0. We do not impute any value for when a story was not shown to a user.

We find that users in the treatment group had a 60\% (S.E. 17\% ) higher total \emph{Total Story Engagement} with content in the \emph{Recommended Stories} section of the app compared to users in the control group. Treated users also completed 78\%  (S.E. 26\%) more stories and spent 87\% (S.E. 25\%) more time consuming content on the personalized section. We document significant patterns of heterogeneity in treatment effects. Users who consume more niche content (i.e., content that is less popular overall) in the pre-experimental period have substantially higher treatment effects than users who liked popular content. This is consistent with the fact that the editorial team selects content targeted to typical -- rather than niche -- tastes. Therefore, users with preferences different from the majority are likely to benefit more from personalized content. Furthermore, users with long histories of content interactions also gained more from content personalization. This is consistent with the nature of matrix factorization, which learns user-specific preferences from historical data. We expect better performance for users with more information about past interactions. Lastly, we find that users who did not have any non-zero \emph{Story Engagement} with any stories in the \emph{Recommended Story} section in the past had a particularly high treatment effect: an 87\% increase in \emph{Total Story Engagement} (S.E. 25\%). The positive treatment effect for users not interacting with stories in this section suggests that users explore the app enough to notice content they prefer even in a section they rarely use and adjust their consumption decisions accordingly.

Users who receive personalized recommendations in the \emph{Recommended Story} section have a higher engagement with the app across all sections. We find a 14\% (S.E. 7\%) increase in the total \emph{Story engagement} on the app, a 19\% (S.E. 8\%) increase in the number of completed stories, and a 20\% (S.E. 9\%) growth in the time spent reading stories. These results suggest that the increased usage of the \emph{Recommended Story} section is not driven entirely by substitution away from other sections of the app. On the contrary,  we find that users substitute away from other non-app activities to use the app more. In summary, better content selection can increase the overall total content engagement of a digital education app, justifying, from the company's perspective, investments in developing recommendation systems.

The rest of the paper is organized as follows. \Cref{lit_reviw} discusses related literature. \Cref{empdata} details the empirical setting. \Cref{sectionoffline} presents the methodology used to develop and test the recommendation model using offline data. \Cref{RCT} describes the design of the randomized experiment and presents the results. Section \ref{conclusion} concludes.

\section{Literature review}\label{lit_reviw} This paper relates to several strands of literature. Personalized recommendation systems have been studied intensively in the entertainment industry \citep{davidson2010youtube, gomez2015netflix,  jacobson2016music, holtz2020engagement} and in the retail shopping industry\citep{linden2003amazon,   sharma2015estimating, smith2017two,greenstein2018personal, ursu2018power}. For example, in the entertainment context and using a similar approach to our paper, \citep{holtz2020engagement} show that personalized recommendations increase the consumption of podcasts on Spotify. However, there is little empirical evidence of the usefulness of recommendation engines beyond entertainment platforms and e-commerce.\footnote{\cite{DBLP:reference/sp/DrachslerVSM15} provide an extensive review of literature on recommendation systems in digital education and point out a shortage of papers documenting the efficiency of recommendation systems using reliable evaluation methods. They conclude the review by calling for more comprehensive user studies in a controlled experimental environment.} Additionally, we show that personalized recommendations can be effectively boost user engagement in settings with moderate amounts of data.

\par The existing evidence of the efficacy of recommendation systems in education is generally based on small studies that combine the introduction of personalized recommendations with other changes to the user interface. \cite{ISIS} use a recruited group of university students to study the effect of showing personalized recommendations of course materials to not showing any recommendations at all. While this study based on a randomized experiment, it bundles two changes in one treatment: adding a user interface element and personalizing recommendations. Furthermore, this study is based on a relatively small sample of 250 subjects. \cite{Ruiz-Iniesta2018} develop and test a recommendation system on a digital education platform called \textit{Smile and Learn}, and evaluate it in an observational study. Their proposed treatment is a new user interface component with recommendations generated using matrix factorization. The newly introduced system helps users navigate the app and reach desired content quicker. They find substantial increases in consumption of recommended items versus non-recommended items. However, the treatment in \cite{Ruiz-Iniesta2018} has two elements: the part simplifying app navigation by adding a user interface component and a personalization component. Our work provides results that isolate the impact of personalization on the consumption of learning items. To the best of our knowledge, our paper is the first large-scale study in the digital education context that estimates the effect of personalization on user engagement in isolation from other changes in the app.

Second, our work contributes to the growing literature assessing the effects of personalized recommendation systems on the diversity of consumed content. To our knowledge, we are the first to do so in a digital education context. \cite{anderson2020algorithmic} and \cite{holtz2020engagement} provide evidence from a randomized experiment indicating that personalized recommendations reduce the diversity of content consumed on \emph{Spotify}. In the context of retail, \cite{AMAZONREC} show that, while recommendations reduce within-consumer diversity, their effect on aggregate diversity is ambiguous. \cite{NEWSREC} find that recommendations reduce consumption diversity in the context of news consumption. In this paper, we show that users with niche preferences are recommended more niche content and less often interact with stories liked by the majority of users.\footnote{This closely relates to the literature documenting 'filter-bubbles' due to the personalization of content on media platforms  \citep{haim2018burst, moller2018not}.}

Lastly, this paper relates to a rich literature on technology-assisted language learning.\footnote{See \cite{garrett2009computer}, \cite{zhao2003recent}, and \cite{tafazoli2019technology} for reviews of this literature.} Personalization in the language learning context has been shown to be effective in task assignment \citep{xie2019personalized} and learning resource recommendations \citep{sun2020vocabulary}. We contribute to this literature by bringing causal evidence of the impact of personalization on time spent interacting with language learning content.

\section{Empirical setting}\label{empdata}
To investigate the impact of personalization on engagement in an educational context, we partnered with \textit{Stones2Milestones (S2M)}. This company operates in India and provides technology-enabled English education through a variety of programs serving a diverse set of users. The main product of \emph{S2M} is a smartphone app called \emph{Freadom}, aimed at 3 to 12-year-old children. In 2021, the app had approximately 7,5000 average daily users. Users come to the app through two main channels: business to business (``B2B''), in particular partnerships with schools, in which schools later recommend the app to their students, and business to consumer (``B2C''), in which independent users download the app from the app store. Additionally, a paid version of the app gives access to some additional non-essential features.

\begin{figure}[!ht]
			\caption{Screenshots from \emph{Freadom}.}\label{screenshot}

			\subfloat[Home Feed page: users open the app on this page.]{\includegraphics[scale = 0.13]{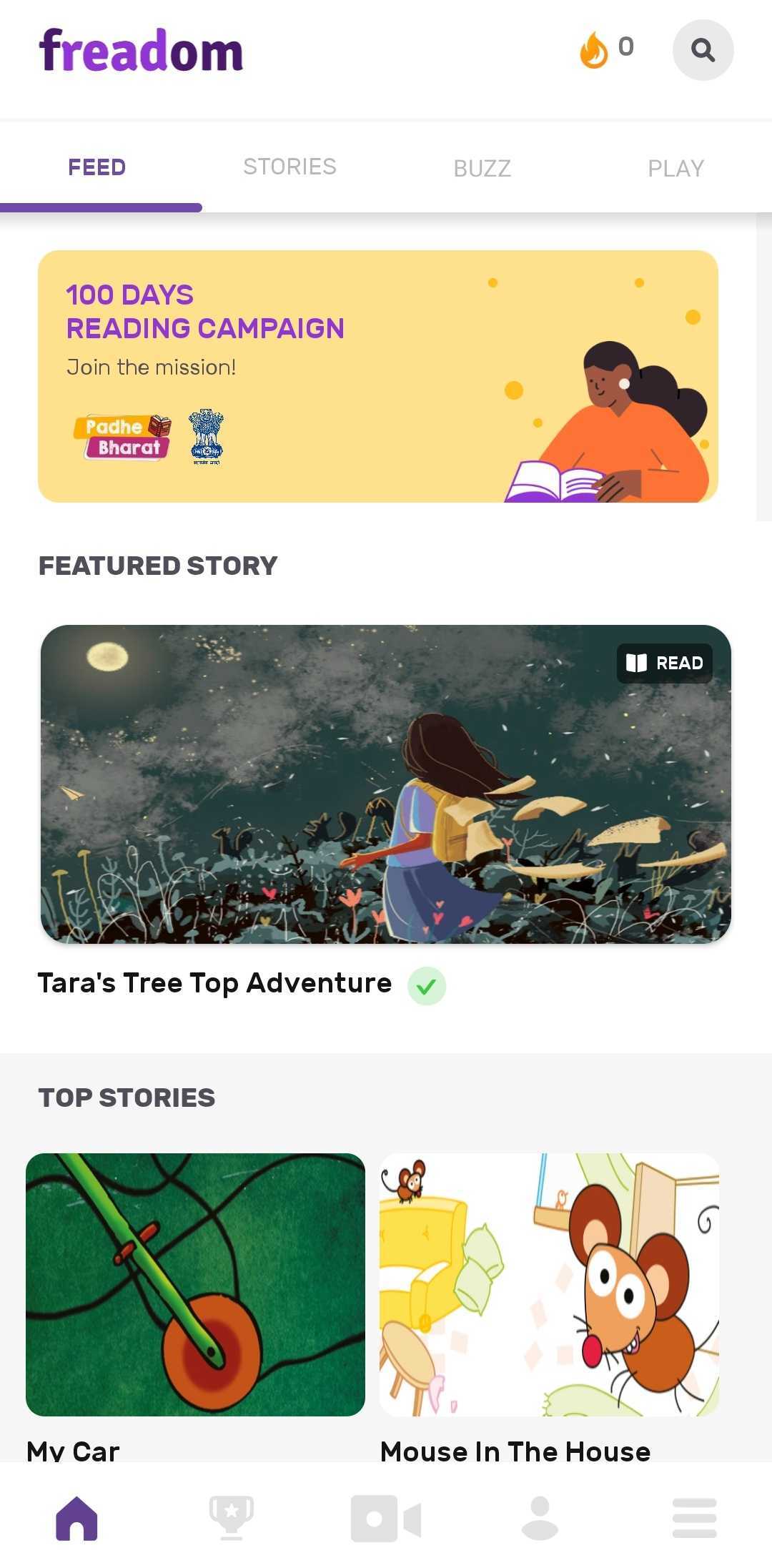}\label{fig:f1}}
			\hfill
			\subfloat[Stories page: contains all story sections.]{\includegraphics[scale = 0.13]{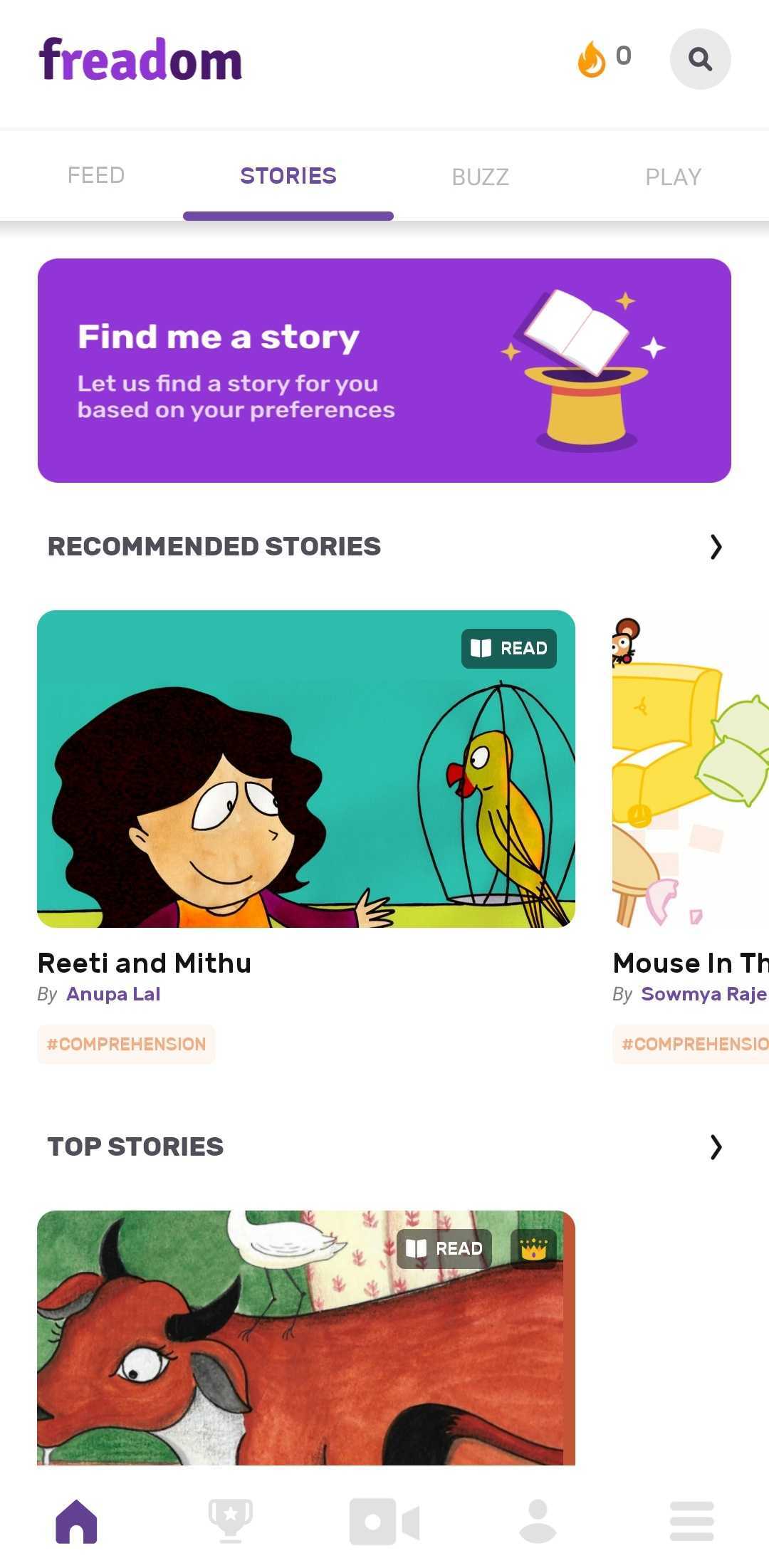}\label{fig:f2}}
			\hfill
			\subfloat[\emph{Recommended Stories}: a section page containing a slate of stories.]{\includegraphics[scale = 0.13]{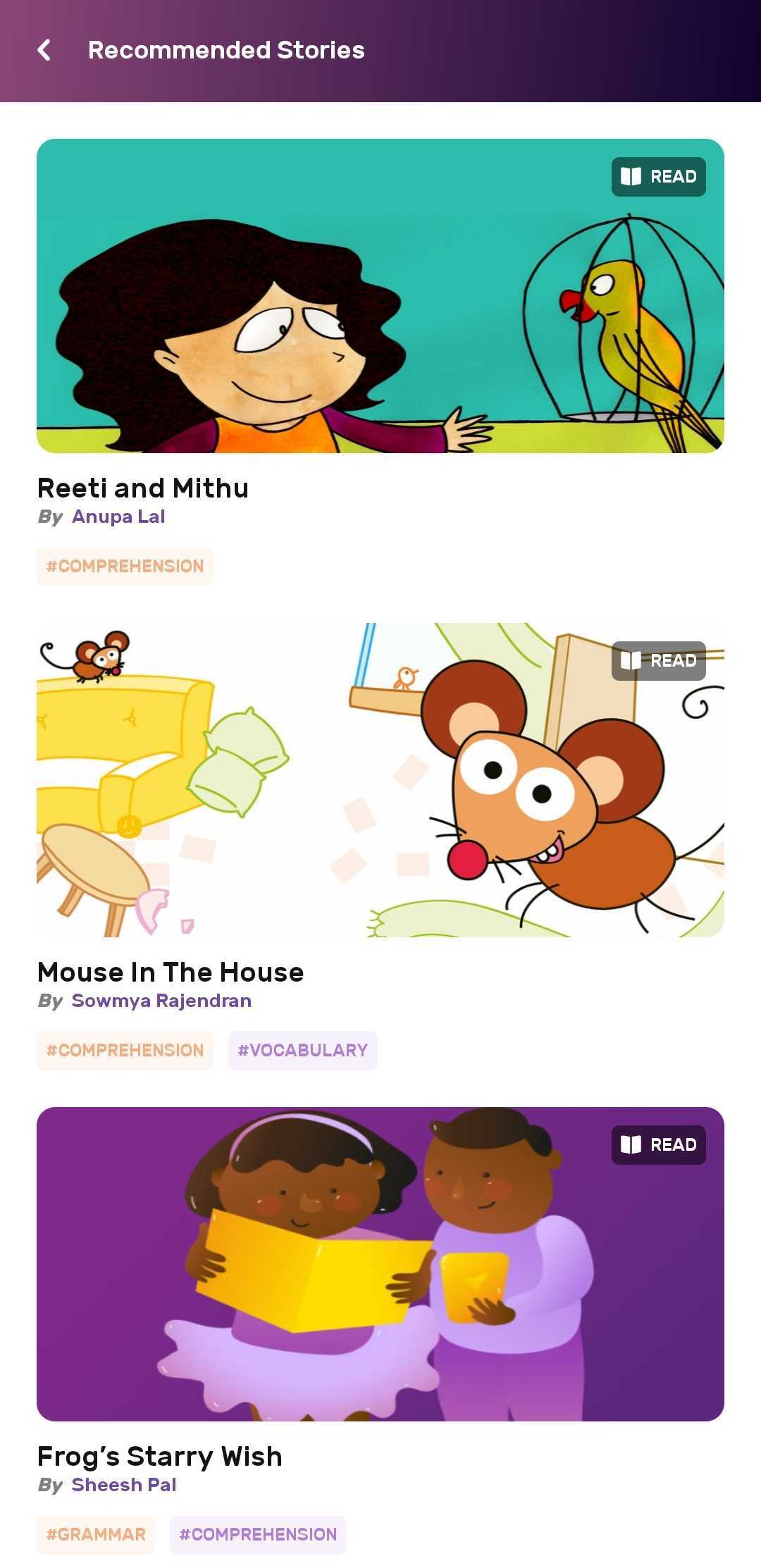}\label{fig:f3}}
		\end{figure}

The main content of \emph{Freadom} is short illustrated stories. Stories are organized in different sections based on various themes such as \emph{Trending now} or \emph{Recommended Story}. Figure \ref{screenshot} shows screenshots from the app. Each section displays a slate of stories that a user can browse and choose the ones to read.\footnote{A slate is a sequence of stories displayed horizontally. A user can swipe left or right in a slate to move between stories.} The selection of stories into sections follows various rule-based algorithms. For example, \emph{Trending now} displays stories that many users currently consume. \Cref{fig:f3} presents the top part of the \emph{Recommended Story} section. Importantly, none of the app sections were personalized during the pre-experimental period.

\emph{Freadom} stories are curated by the \emph{S2M} pedagogical team and  publishers specializing in educational content for kids. They are age-appropriate and created with an educational goal in mind. As such,\emph{S2M} operates under the premise that maximizing the consumption of content on the app helps learners achieve their educational goals.

\begin{figure}[H]
			\centering
			\caption{User-story interaction engagement funnel}
			\includegraphics[scale = 0.60]{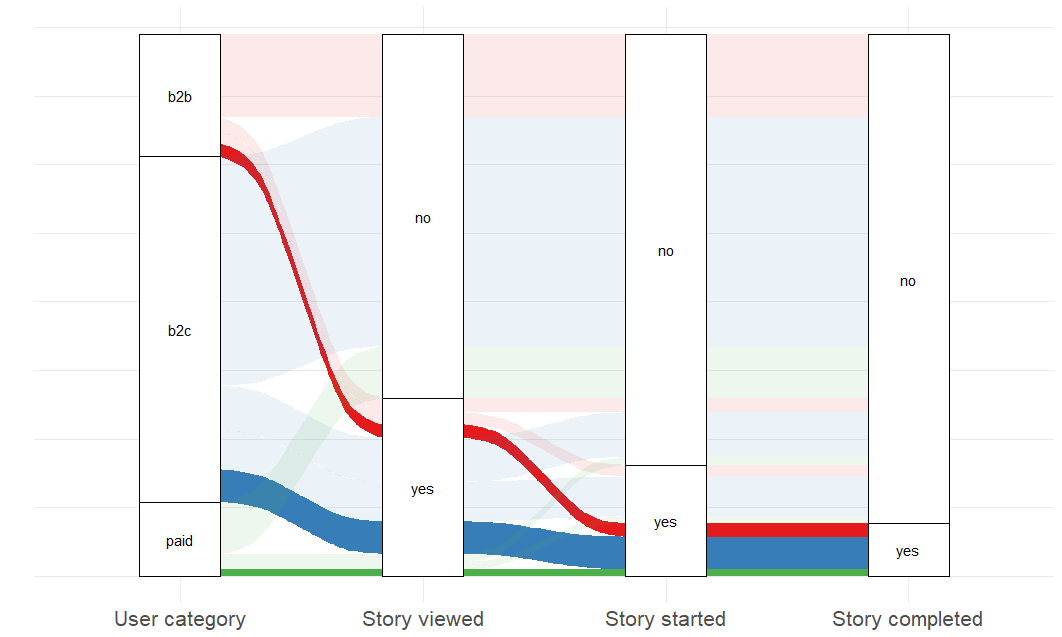}
			\caption*{\footnotesize{\textit{Note: Story Engagement funnel broken by the type of user (B2B, B2C, paid) and the outcome of user-story interactions. The highlighted colors show the share of user-story interactions that resulted in story completion and background colors interactions that did not lead to story completion. In red are B2B users, in blue are B2C, and in green are paid users. The first column shows shares of user categories, the second column presents the share of users that viewed the story, the third the share of users that started the story, and the fourth that completed it.}}}
			\label{fig:utility_funnel}
		\end{figure}

\emph{Freadom} users browsing in a section can click on stories, which will display their descriptions. Many users who view the description decide to go back to browsing, while others start reading the story. Only a minority of user-story interactions lead a user to read a story completely. \Cref{fig:utility_funnel} shows a content interaction funnel that represents the frequencies of users' content consumption decisions. We divide users into three main categories: \emph{B2C}, \emph{B2B}, and \emph{paid} and show frequencies of different outcomes from interactions with stories. The interaction between a user and a story is the unit of interest. Users decide whether to view a story or not (second column), whether to start reading it (third column), and whether to complete it or not (final column). We find that users tend to explore many stories and acquire information about them by viewing or starting before deciding which stories to read.

\section{Using offline data to develop a recommendation system}\label{sectionoffline}
The recommendation system we propose is based on several components. First, an outcome model predicts user engagement with a specific story. Predictions from the outcome model are used to rank stories for a specific user, and stories with the highest predicted engagement are ranked toward the top. Second, we define a target audience since a specific recommendation system might perform better for a subset of users. As such, a selected group of users receives recommendations from the proposed model. Finally, the re-ranking routine determines how the selection of stories changes over time as users engage with them. 

\subsection{Conceptual framework and outcome models}

Consider users indexed by $i$, where $i \in \mathcal{I} \equiv 1,...,I$ and stories indexed by $j$, where $j \in \mathcal{J} \equiv 1,...,J$. Let $Y_i(j)$ be the potential outcome for user $i$ when exposed to story $j$, a quantity that is defined for all $i$ and $j$. That is if user $i$ was (counterfactually) exposed to story $j$, $Y_i(j)$ represents the outcome that would occur. In the observed data, users are only exposed to a subset of stories, denoted $\mathcal{J}_i$. Let $U_{ij}$ be the observed value of the outcome when $i$ is exposed to $j$. Let $\sigma(x) = 1/(1 + e^{-x})$, the sigmoid function. 

\begin{figure}[H]
	\centering
	\caption{Users and slates}
	\includegraphics[scale = 0.70]{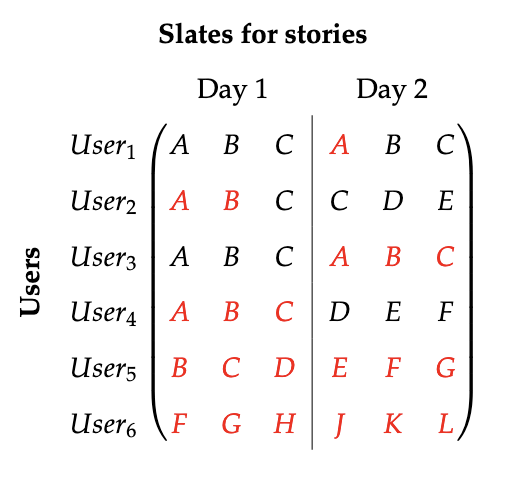}
	\caption*{\footnotesize{\textit{Note: Visualization of the organization of users and slates of their stories. In \textbf{black} are the stories assigned to a user by a recommendation system. In \textbf{red} are those that a user engaged with. Low-engagement users are at the top (users 1 and 2), and high-engagement users are at the bottom.}}}
	\label{fig:slates_1}
\end{figure}

Users engage with stories that are assigned to them. The assigned stories are organized in slates that are specific to each user. In Figure \ref{fig:slates_1}, we show an example of a matrix of user-story assignments. We show 6 users and consider slates of 3 stories. In black, we present all stories the recommendation system selected for users, and in red, stories that a user engaged with. In \cref{fig:slates_1}, we consider a case of an editorial-based system in which all users initially get the same selection of stories. However, when a user engages with a story, this story is removed from slates prepared for subsequent days. Generally, high-engagement users will see more stories in total than low-engagement users because the stories they engaged with in the past have been removed. Consequently, stories in the lower rank on a slate will be engaged by only heavy engagement users. In Figure \cref{fig:slates_1} we order rows by the overall level of engagement of a user presented in that row. Low-engagement users are at the top (users 1 and 2), while users towards the bottom are high-engagement users (users 5 and 6).


The selection of stories into slates is based on a prediction of an outcome model. Consider the popularity-based model (TWFE), which estimates $\mathbb{E}[U_{ij}]$ by fitting observed outcomes to the model. Consider:

\begin{equation}\label{eq:twfe_model}
\mathbb{E}[U_{ij}]=\sigma(\beta_{0} + \Psi_{i} + \Gamma_{j}),\end{equation}

where $\Gamma_{j}$ and $\Psi_{i}$ are user and story fixed effects, respectively. The TWFE outcome model leads to a non-personalized recommendation system, which means that for a given user, stories are simply ranked by their popularity across all users. Note further that this model can only be considered an approximation because of the discrete support of the outcome we consider. The TWFE model can be a useful benchmark for evaluating personalized models because it is easy to implement. It is useful to show that simpler models do not achieve similar performance as the personalized model to justify the development and introduction of a more complicated personalized model.

In the historical data, assignments of users to stories were not personalized. As a consequence, if each user was exposed to a single story $J_i$ per unit of time, then treatment assignment would be independent of potential outcomes ($Y_i(j)\indep \mathcal{J}_i|$), and (letting the number of observations for each story and user grow large) the TWFE would be consistent of $\mathbb{E}[Y_i(j)].$ However, in practice, some users use the app more, which affects the selection of stories they see. These users may also have higher engagement per story. The inclusion of user-fixed effects adjusts for these differences, but the consistency of the estimates depends on the additive separability assumption embedded in the TWFE model. 

Another option to consider in lieu of the TWFE model is the matrix factorization approach \citep{rendle2010factorization} which uses the following outcome model:
\begin{equation}\label{eq:cf}
    \mathbb{E}[U_{ij}] = \sigma(\Lambda_{j} \times \Theta_{i} + \beta_{0} + \Psi_{i} + \Gamma_{j}),
\end{equation}
where $\Lambda_{j}$ is a latent preference vector per user and $\Theta_{i}$ a latent vector per story, where each vector has length $k$. A given component of the story vector can be interpreted as a latent characteristic (e.g., a serious or a funny story), and the corresponding user component can be interpreted as the preference for that characteristic. This matrix factorization model identifies a low-dimensional representation of both users and stories so that users with preferences for a particular type of story are located close (in the sense of Euclidean distance) to one another and their preferred story types. 

The matrix factorization model performs well if the matrices $\Lambda$ and $\Theta$ represent underlying preferences well. The more data we have on users' and stories' past interactions, the higher the chance of arriving at an accurate representation of the outcome matrix. Crucially, this depends on the underlying true outcome model: the model performs well if a low-rank matrix well approximates the true matrix and if there is sufficient data (overlapping user and story interactions) to estimate the low-rank approximation. In the next section, we evaluate the performance of an outcome model based on matrix factorization and compare it to the performance of the TWFE model.

\subsection{Selection of an outcome model}

\paragraph{Data.} The analysis in this step relies on historical data consisting of the logs of users' interactions with stories.\footnote{User accounts are verified using a phone number and are unique.} Every entry in this dataset records an interaction of a user with a story, as well as to what extent they consumed the story; specifically, whether a user did not consume it at all, considered reading it by viewing the story description card, started reading it, or completed reading it. The dataset includes a few observables about each user, including information about a user's grade level, as well as observable characteristics of stories, including a tag recording the collection (for example, "animal," or "sport") to which a story belongs.

\paragraph{Constructing an Outcome.} Outcome measures that represent beneficial outcomes of user-story interactions can be constructed in a variety of ways. The ultimate goal of personalization is to increase student learning. While we do not directly observe learning, we proxy for learning through various engagement measures. In consultation with \emph{Stones2Milestones}, we defined the metric \emph{Story Engagement} as follows:
\begin{itemize}
    \item $1$: The user completed the story
    \item $0.5$: The user started but did not complete the story 
    \item $0.3$: The user viewed the story page but did not start the story
    \item $0$; The user was shown the story, but the user skipped it and viewed a different story during the session
    \item $N/A$: The user did not interact with any story shown to them during the session, or the user was not shown a story

\end{itemize}
We distinguish between the user choosing not to engage with a story that was shown (0) and the user never having an opportunity to interact with a story because it was not shown (N/A). This is a critical distinction for understanding user preferences that is not always available in historical datasets.\footnote{Not all apps or websites log what a user sees while clicking on an item typically generates a logged event because it retrieves the information to display to a user.}
We use the \emph{Story Engagement} measure to construct a matrix of outcomes, with users in rows and stories in columns. This is the main building block of the recommendation system.
		
\paragraph{Choice of an outcome model.} We select an outcome model based on predictive accuracy as measured by the Mean Squared Error (MSE):
\[
MSE(m) = \frac{1}{n}\sum_{k=1}^{K}\left(U_{k} -\hat{U}_{k}(m)\right)^{2},
\]
where $m$ denotes the model we are evaluating, $k = 1,..,K$ indexes user-story interactions, $U_{k}$ is the observed outcome from interaction $k$, and $\hat{U}_{k}(m)$ is the predicted outcome of interaction $k$ using model $m$.

\begin{figure}[H]
	\centering
	\caption{Evaluation of an outcome model}
	\includegraphics[scale = 0.60]{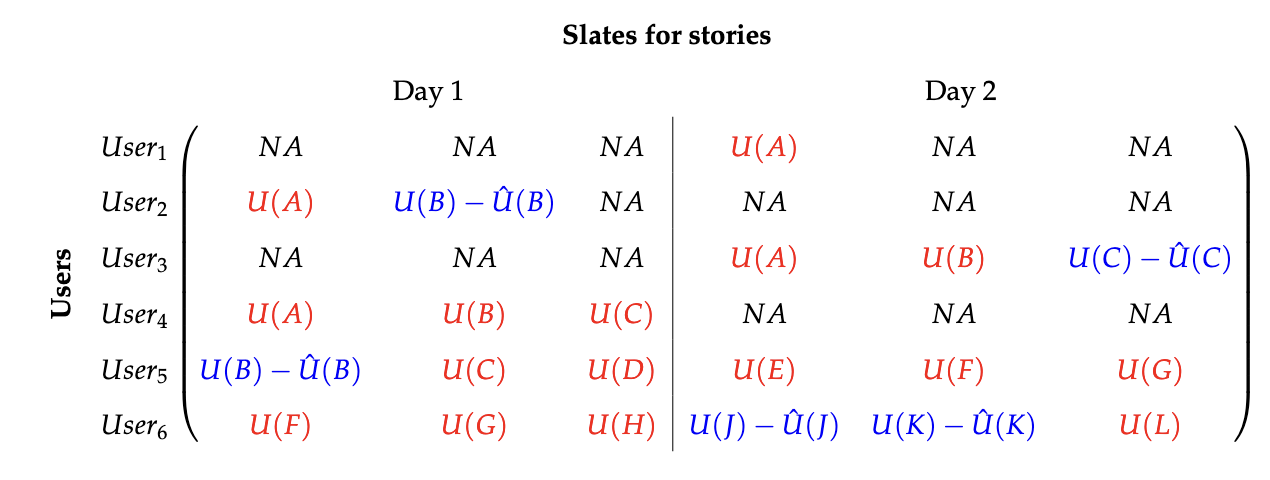}
	\caption*{\footnotesize{\textit{Note: Visualization of the model evaluation. In \textbf{red} we show outcomes of user-item interactions that were used for model training and in \textbf{blue} for model evaluation. $U(\cdot)$ denotes the observed outcome and $\hat{U}(\cdot)$ model's prediction.}}}
	\label{fig:slates_mse}
\end{figure}

We split the historical observational data into training and evaluation sets to evaluate the model's performance. We estimate the parameters of the alternative models on the training set and compare performance on the evaluation set. Figure \ref{fig:slates_mse} visualizes this approach. In red, we show data points used for model training, and in blue for model evaluation. $U(\cdot)$ denotes the observed outcome and $\hat{U}(\cdot)$ model's prediction. Heavy users (users 5 and six) are presented at the bottom of the figure. Since they generate more data points, the model evaluation is mostly based on the model's performance for these users.


\paragraph{Outcome model estimation details.}
The matrix factorization model parameters are estimated using the PyTorch  framework in python ( see \citet{pytorch}). We use Stochastic Gradient Descent (SGD) with the Adam Optimization method. To regularize, we use an L2 penalty on our parameters (that is, we penalize the sum of squares of the parameters). We tune the number of latents, $k$ (the dimension of $\Lambda_{j}$ and $\Theta_{i}$), and our L2 penalty parameter using a randomly held out validation set.\footnote{We split our dataset into a train, test, and validation set at random. Another approach is to split the data by time into a train dataset and a test dataset; so that we test in the period following our training data. The train data is randomly split into a train set and a validation set in this setting. We also executed this approach; this leads to similar results.} Once we have our optimal learning hyperparameters, we relearn our model on the entire dataset, resulting in the final model.

\paragraph{Performance on historical data.}
To test the accuracy of the prediction models, we compare the performance of the popularity-based model from \Cref{eq:twfe_model} to the performance of the matrix factorization model from \Cref{eq:cf}; additionally. For completeness, we include the performance of a model with just a constant term (mean model). \Cref{tab:mse-table_1new} presents the results.

\begin{table}
\caption{MSE values for matrix factorization (PYTF), two-way fixed effects (TWFE), and a simple mean model.}
\begin{center}
\resizebox{0.33\textwidth}{!}{%
\begin{tabular}{@{}rrr@{}}
\toprule
\textbf{PYTF} & \textbf{TWFE} & \textbf{Mean Model} \\ \midrule
0.0962        & 0.1022        & 0.1309              \\ \bottomrule
\end{tabular}%
}
\label{tab:mse-table_1new}
\caption*{\footnotesize{\textit{Note: Models are trained and evaluated on the dataset including all users and stories (i.e., there is no filtering based on user-item history length).}}}
\end{center}
\end{table}

\paragraph{Determining the target audience.}
The performance of the matrix factorization model depends on the length of histories of interactions between users and stories.\footnote{Our approach is generally not suitable for new users and new stories. The so-called cold-start problem of assigning content recommendations to users that have not yet revealed preferences from content interactions or stories whose latent style is still unknown is well-documented in recommendation systems literature, see e.g., \cite{lam2008addressing, lika2014facing, bobadilla2012collaborative}. } To determine the right set of users and stories for the deployment of the recommendation model, we compared the MSEs of \emph{Story Engagement} predictions from the selected outcome model trained over different amounts of data and tested on a held-out evaluation set. The training sets differ by the minimum histories of interactions between stories and users. This analysis tells us how much user and story history is necessary for the recommendation model to provide high-quality recommendations.

\Cref{tab:filters} presents MSEs for nine specifications depending on the length of the history of stories (columns) and of users (rows). We evaluate all specifications on the same dataset with thresholds of 20 interactions for stories and 20 for users.

\begin{table}
\caption{Matrix Factorization Model Mean Squared Error for various user and story histories.}
\begin{center}
\resizebox{0.4\textwidth}{!}{%
\begin{tabular}{@{}crrrr@{}}\toprule
                                     &       \multicolumn{1}{c}{}              & \multicolumn{3}{c}{\textbf{Stories}}                                                  \\
                      & \textbf{}                         & \textbf{20} & \textbf{60} & \textbf{100} \\ \midrule
\multicolumn{1}{c|}{\multirow{3}{*}{\rotatebox[origin=c]{90}{\textbf{Users}}}} & \multicolumn{1}{r|}{\textbf{20}}  & 0.0967      & 0.0931      & 0.0932       \\
\multicolumn{1}{c|}{}                       & \multicolumn{1}{r|}{\textbf{60}}  & 0.0964      & 0.0931      & 0.0931       \\
\multicolumn{1}{c|}{}                       & \multicolumn{1}{r|}{\textbf{100}} & 0.0959      & 0.0930      & 0.0931      \\ \bottomrule   
\end{tabular}%

}
\caption*{\footnotesize{\textit{Note: The rows represent the minimum interactions per user, and the columns represent the minimum interactions per story. We use a single trained model on the largest dataset (20, 20) and report MSEs on different test sets.}}}
\label{tab:filters}

\end{center}
\end{table}

Based on the results from \Cref{tab:filters}, we selected the population of users that would receive personal recommendations to be the set of users and stories with at least 60 interactions in our historical data. Two factors contributed to this decision; first, high-quality predictions as measured by MSE and, second, the sample size requirement for the randomized experiment.\footnote{Approximately 15\% of users in the entire user base and 92\% of all stories in the app have at least 60 interactions.}

\paragraph{Selecting a section of the app.}
\emph{Freadom} is built based on multiple scrollable sections. Sections vary by popularity, and one important driver of a section's popularity is its position on the page. The most popular sections on the app are \emph{Popular}, \emph{Trending Now}, \emph{Recommended Story}, and \emph{Today For You}.

We chose to deploy the recommendation model in the \emph{Recommended Story} section. This section was popular among more experienced users of the app, which meant that many users of this section met the 60 interactions threshold. Additionally,  \emph{S2M} had also initially intended the section to be for personalized recommendations hence the name - \emph{Recommended Story}.

\paragraph{Re-ranking over time.}
The \emph{Recommended Story} section presents stories in a slate of 15 entries. The slate design task involves deciding how to rank the stories and how frequently to update the ranking. We wanted to keep the ranking and refreshing module similar to the baseline so we could focus on isolating the effects of personalization.

Due to computational constraints, new \emph{Story Engagement} predictions were generated once per week. Thus, the algorithm ranked stories in decreasing order of predicted outcome each week, and the top 15 stories were assigned to the slate for the next week. On every day during the week, the algorithm removed the stories that were started and were replaced by stories next in the predicted \emph{Story Engagement} ranking.

\Cref{alg:rec_system} summarizes the functioning of a recommendation system. $L_{D}$ is the data, and it consists of the history of interaction results $y_{ij}$ until day $D$. Slate $s_{i,D'}$ is a sequence of 15 stories to be displayed to a user $i$ during the day $D'$. The recommendation system takes the predicted \emph{Story Engagement}, $\hat{U}_{ij}$, from model$m$ for all stories that a user $i$ can engage with and selects 15 top stories. After Day 1, the \emph{Ranking Algorithm} updates slates until the end of the week: Whenever a user was active in the section for two days but did not complete any of the top 3 stories, we removed those stories from the slate shown to the user. Whenever a user is active in the app, they see the top of the slate without leaving a record that they have interacted with these stories. We use this fact to infer that the user did not like the top 3 stories. If the user has been inactive in the section, on the other hand, the ranking stays the same.

\begin{algorithm}
\caption{Recommendation System}\label{alg:rec_system}
\begin{algorithmic}
    \State {\bfseries Input:} Data $L_{D}$, outcome model $m$ 
    \State {\bfseries Output:} Slates $s_{i,D'}$, $\forall i$ and $D' \in \left\{D + 1, D + 2, D + 3, D+ 4, D+5, D+6, D+7\right\}     $
  \end{algorithmic}
  
\begin{algorithmic}[1]
    \If{$t = 1$}
      \State $U_{i,D} = m(L_{D}, i)$       \Comment{Get full recommendations}
      \State $s_{i,D+t} = \text{top\_k}(U_{i,D}, 15)$
      \EndIf

    \If{$t >= 2$ and $t <= 7$}
        \State $s_{i,D+t} = \text{top\_k}(U_{i,D} - \text{started\_stories}(L_{D+t-1}), 15)$  \Comment{Remove stories started}
        \If{$\text{active\_user}(i)$} \Comment{If user isn't active, no update}
          \If{$\text{completed\_stories}(L_{D+t-1}) \cap \text{top\_k}(s_{i,D}, 3) = \emptyset$} \Comment{If user didn't complete any of top 3 stories}
          \State $s_{i,D+t} = \text{top\_k}(U_{i,D} - \text{top\_k}(s_{i,D}, 3), 15)$ \Comment{Remove the top 3 stories}
          \EndIf
        \EndIf
    \EndIf
    
  \end{algorithmic}

\end{algorithm}

 Function $top\_k(\cdot)$ selects top $k$ stories based on the predicted \emph{Story Engagement}.

\section{Recommendation Policy Evaluation in a Randomized Controlled Trial}\label{RCT}
To evaluate the selected recommendation system, we carry out a randomized experiment. The objective is to compare \emph{Total Story Engagement} under the existing editorial-based system and the personalized recommendation based on the matrix factorization approach.

\subsection{Design of the experiment}

In the experiment, 7750 users were randomized into treatment and control. We only consider users who had at least sixty story interactions before the experiment. The treatment group received personalized recommendations in the \emph{Recommended Story} section, and the control group received stories from the existing editorial-based system. The section's user interface was consistent across the control and treatment groups; the only thing exogenously varied was the set of stories displayed in the section. The content presented in the other sections of the app was left unchanged, and treated users were not aware of the change in the recommendation system. Based on discussions with the partner, we decided to run the experiment for two weeks. Based on the analysis of past data, the minimum detectable effect on \emph{Total Story Engagement} was 0.08 standard deviations.

The experiment started on July 22,  2021 and lasted until August 4.\footnote{After the experiment, the personalized recommendations system was launched for all eligible users on the \emph{Recommended Story} section.} During the experiment, 3023 users from the experimental groups launched the app at least once, and of them, 525 viewed at least one story in \emph{Recommended Story} section.\footnote{The significant difference in the number of randomized students and the number of students who were active during the experimental period is because (i) many students were only active in other sections, and (ii) there is continuous churn as students drop off the app over time.} We report the balance of observable characteristics between the treatment and control groups in Appendix \ref{cobal}.  

In the evaluation of the experiment, we consider subjects who launched the app at least once during the experiment. This means that we exclude users who did not launch the app during the experiment period, but we include users who launched the app but did not click on any of the stories in the \emph{Recommended Story} section. The reason for including the latter group is that users can see the front page of the first story in \emph{Recommended Story} section without starting to interact with any of the stories in the section, so we also capture the change from not interacting with content in the \emph{Recommended Story} to having some non-zero story engagement interaction.

\subsection{Outcome metrics}

We focus on two types of outcomes: outcomes specific to \emph{Recommended Story} section and outcomes measuring the overall app usage. Even though other sections in the app remained unchanged, we are also interested in the impact on overall app usage to understand whether changes in one section affect the utilization of content elsewhere or whether the overall time spent on the app shifts. In this specific context, where many users are consuming content based on the recommendation of parents or teachers, understanding the overall elasticity of consumption with respect to changes in app quality is an important strategic metric that can guide app development.

We consider the following outcome metrics: (i) \emph{Total Story Engagement} -- per user sums of Story Engagement from all user-story interactions in \emph{Recommended Story} section during the experiment, (ii) \emph{Total Story Engagement all sections} -- per user sums of Story Engagement from all user-story interactions in all sections of the app during the experiment, (iii) \emph{total stories} -- per user sums of completed stories in \emph{Recommended Story} section during the experiment, (iv) \emph{total stories all sections} -- per user sums of completed stories in all sections, (v) \emph{total reading time} -- per user sums of estimated reading time of stories completed in \emph{Recommended Story} section, and (vi) \emph{total reading time all sections} -- per user sums of estimated reading time of stories completed in all sections.\footnote{The estimates of the reading time per story are provided by \emph{S2M} as intervals, e.g., from two to four minutes. For each story, we take the midpoint of the interval.}

We constructed all variables based on raw log files provided by \emph{S2M}. These log files are internal data used by \emph{S2M} data analytics teams; they constitute the most accurate available picture of users' behavior on the platform. Nevertheless, occasional instrumentation errors occur. The type of instrumentation error that is problematic for our analysis is an incorrect attribution of user-story interactions, which can take a form of a user being assigned interactions of another user or assigned completions instead of views. This results in some users having spurious, very high utilization during specific sessions. To avoid including such sessions in the analysis, we drop users -- 40 in total -- with at least one session in which they completed more than ten stories. \Cref{tab:sum_stats_exp} provides summary statistics of variables describing utilization in the \emph{Recommended Stories} section.

\begin{table}[!htbp] \centering 
  \caption{Summary statistics of outcomes on the \emph{Recommended Stories} section per group.} 
  \label{tab:sum_stats_exp} 
  \resizebox{\textwidth}{!}{
\begin{tabular}{>{}l|lrrrrrr}
\toprule
names & group & min & mean & percentile 75th & percentile 90th & percentile 95th & max\\
\midrule
 {\textcolor{black}{\textbf{Total Story Engagement}}} & control & 0 & 0.28 & 0 & 0.51 & 1.6 & 13.6\\
 {\textcolor{black}{\textbf{Total Story Engagement}}} & treatment & 0 & 0.45 & 0 & 1.30 & 3.0 & 23.9\\
 {\textcolor{black}{\textbf{Total stories}}} & control & 0 & 0.15 & 0 & 0.00 & 1.0 & 11.0\\
 {\textcolor{black}{\textbf{Total stories}}} & treatment & 0 & 0.27 & 0 & 1.00 & 2.0 & 21.0\\
 {\textcolor{black}{\textbf{Total reading time}}} & control & 0 & 1.04 & 0 & 0.00 & 7.5 & 78.5\\
 {\textcolor{black}{\textbf{Total reading time}}} & treatment & 0 & 1.94 & 0 & 7.25 & 11.0 & 148.5\\
\bottomrule
\end{tabular}
}
\caption*{\footnotesize{\textit{Note: Summary statistics of variables measuring utilization of the \emph{Recommended Story} section during the experiment. The sample includes only users that launched the app during the experiment period. }}}
\end{table}
   
Even though we only consider users who launched the app during the experiment, most of them had zero utilization of the app in the \emph{Recommended Story} section. Nevertheless, we still include them in the experiment evaluation because different recommendation policies might impact the share of users who consume any content in the section. As shown in \Cref{tab:sum_stats_exp}, the treatment group has higher mean utilization and higher utilization at the 90th and 95th percentiles. 

We are also interested in the impact of the personalization of content recommendations in \emph{Recommended Stories} section on the overall app usage. \Cref{tab:sum_stats_exp_allpaths} presents summary statistics of variables describing utilization on all sections in the app.

\begin{table}[!htbp] \centering 
  \caption{Summary statistics of outcome variables on all sections per group.} 
  \label{tab:sum_stats_exp_allpaths} 
  \resizebox{\textwidth}{!}{
\begin{tabular}{>{}l|lrrrrrr}
\toprule
names & group & min  & mean & percentile 75th & percentile 90th & percentile 95th & max\\
\midrule
 {\textcolor{black}{\textbf{Total Story Engagement}}} & control & 0 & 3.79 & 4.47 & 11.4 & 17.77 & 81.5\\
 {\textcolor{black}{\textbf{Total Story Engagement}}} & treatment & 0 & 4.32 & 5.50 & 13.5 & 19.02 & 63.7\\
 {\textcolor{black}{\textbf{Total stories}}} & control & 0 & 1.89 & 2.00 & 6.0 & 9.00 & 41.0\\
 {\textcolor{black}{\textbf{Total stories}}} & treatment & 0 & 2.25 & 3.00 & 7.0 & 11.00 & 42.0\\
 {\textcolor{black}{\textbf{Total reading time}}} & control & 0 & 12.65 & 14.50 & 40.0 & 65.32 & 273.0\\
 {\textcolor{black}{\textbf{Total reading time}}} & treatment & 0 & 15.15 & 15.00 & 46.0 & 73.12 & 365.0\\
\bottomrule
\end{tabular}
}
\caption*{\footnotesize{\textit{Note: Summary statistics of variables measuring the overall app utilization during the experiment. The sample includes only users that launched the app during the experiment period. }}}
\end{table}

\Cref{tab:sum_stats_exp_allpaths} shows that mean outcomes are higher in the treatment group for all outcome variables. Treatment has greater or equal effects at the 75th, 90th, and 95th percentiles.

\begin{figure}%
    \centering
    \caption{Distribution of Total Story Engagement in \emph{Recommended Stories} section per group}%
    \subfloat[\centering Cumulative distribution function per group. Treatment in blue, control in red.]{{\includegraphics[width=15cm]{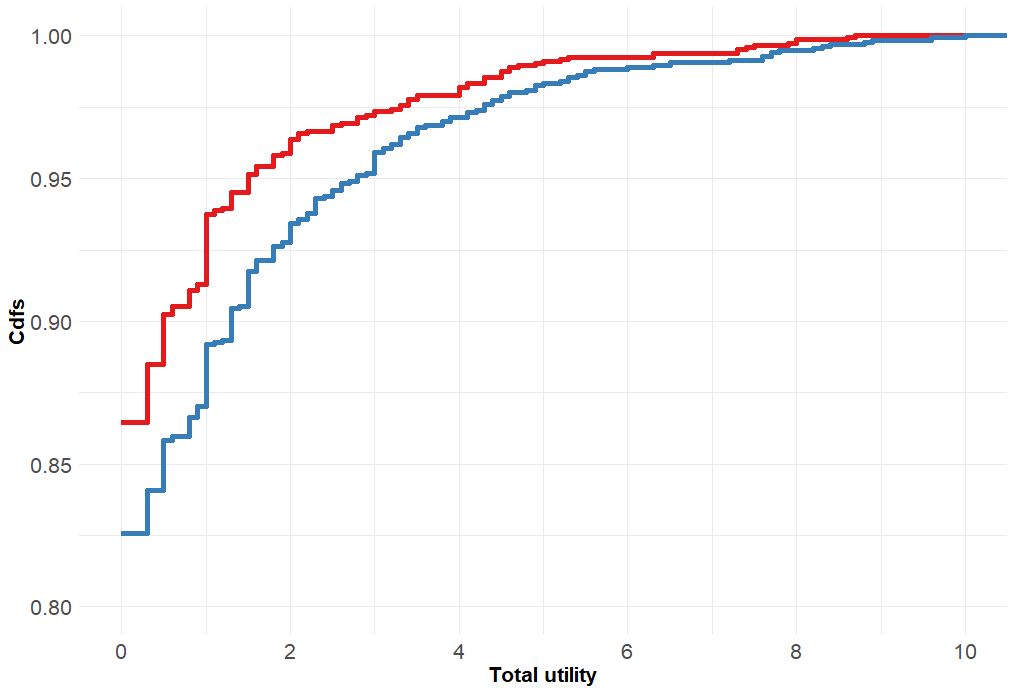} }}%
    \qquad
    \subfloat[\centering Difference between the probability density functions of treatment and control groups. Treatment in blue, control in red. ]{{\includegraphics[width=15cm]{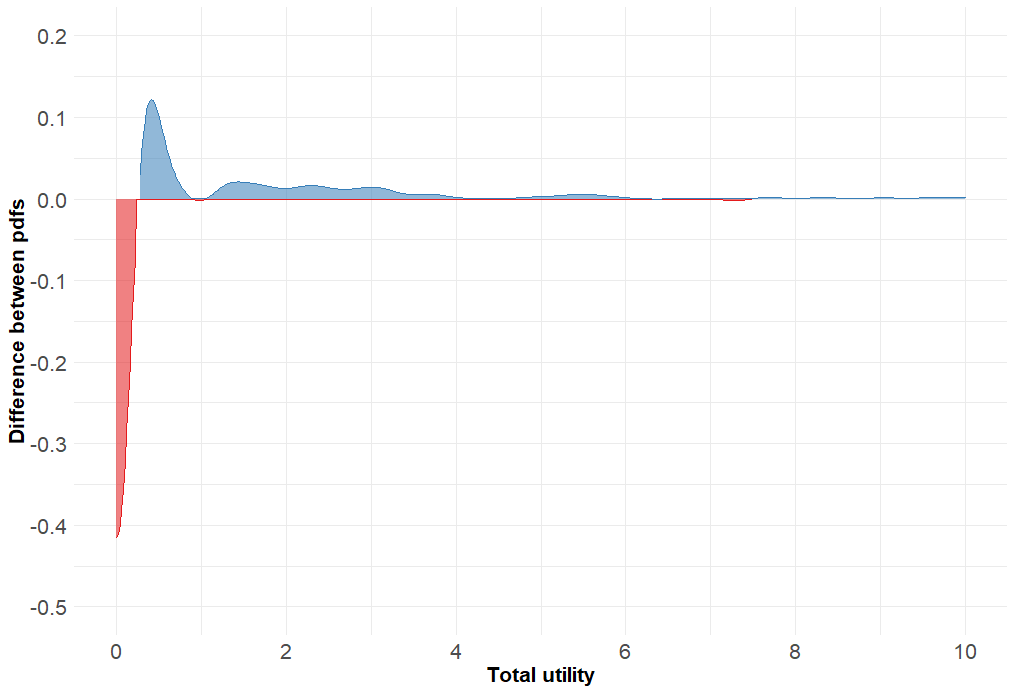} }}%

    \label{fig:total_utility_distribution}%
\end{figure}

To compare distributions of \emph{Total Story Engagement} in the treatment and control groups, we carry the out the one-sided alternative of the Wilcox test. For the \emph{Total Story Engagement} in the \emph{Recommended Story} section, we reject the hypothesis that the true location shift is less than zero, with a p-value of 0.0007. For \emph{Total Story Engagement} on all sections in the app, we can reject the null hypothesis with a p-value of 0.05.
In  \Cref{fig:total_utility_distribution}, we present the entire distributions of \emph{Total Story Engagement}. Panel A shows cumulative distribution functions of \emph{Total Story Engagement} from \emph{Recommended Story} section per experimental group. Panel B shows the difference between the treatment's and control group's probability density functions. These figures show that a larger share of control group users did not have any positive-utility content interaction during the experiment. The treatment group has a higher probability mass for almost any non-zero story engagement.

\subsection{Average treatment effects}\label{ATE_section}
Estimates of the average treatment effects are presented in \Cref{tab:ATE}. We consider three estimators: the difference in mean outcomes between treated and control, a linear regression adjusting for observed user characteristics, and the augmented inverse propensity weighing (AIPW) estimators. The latter two estimators can account for any imbalance in characteristics between treated and control groups and also can potentially reduce the variance of estimators. However, in this dataset, the adjustments do not make a noticeable difference.

We find a strong positive effect of personalization on all outcome variables. The impact on utilization of the \emph{Recommended Stories} section has high economic and statistical significance. \emph{Total Story Engagement} increases by 60\% (s.e. 17\%), the number of stories completed in the section by 78\% (s.e. 26\%), and total reading time by 87\% (s.e. 25\%).
We also find an increase in the utilization of the app across all sections: \emph{Total Story Engagement}  increases by 14\% (s.e. 7\%), the number of stories completed by 19\% (s.e. 8\%), and the reading time in all sections by 20\% (s.e. 9\%). Thus, the increase in content consumption in \emph{Recommended Stories} did not come entirely at the expense of consumption in other sections-- on the contrary, this evidence suggests that users started using the app more.\footnote{In \Cref{ap_out}, we provide a robustness check of these estimates by trimming the top 5\% users with the highest daily number of completed stories instead of the cap on ten stories.}

\begin{table}[!htbp] \centering 
  \caption{Estimates of average treatment effects} 
  \label{tab:ATE} 
\resizebox{\textwidth}{!}{%
\begin{tabular}{>{}l|rrrrrrrr}
\toprule
variable & ATE & S.E. & p-value & ATE \% & ATE reg adj. & S.E. reg adj. & ATE AIPW adj. & S.E. AIPW adj.\\
\midrule
 {\textcolor{black}{\textbf{Total Story Engagement RS}}} & 0.17 & 0.05 & <0.01 & 60 & 0.18 & 0.05 & 0.18 & 0.05\\
 {\textcolor{black}{\textbf{Total stories RS}}} & 0.12 & 0.04 & <0.01 & 78 & 0.13 & 0.03 & 0.13 & 0.04\\
 {\textcolor{black}{\textbf{Total reading time RS}}} & 0.90 & 0.26 & <0.01 & 87 & 0.96 & 0.25 & 0.98 & 0.25\\
\addlinespace
 {\textcolor{black}{\textbf{Total Story Engagement all sections}}} & 0.52 & 0.27 & 0.05 & 14 & 0.50 & 0.26 & 0.51 & 0.26\\
 {\textcolor{black}{\textbf{Total stories all sections}}} & 0.36 & 0.16 & 0.03 & 19 & 0.36 & 0.15 & 0.35 & 0.15\\
 {\textcolor{black}{\textbf{Total reading time all sections}}} & 2.50 & 1.10 & 0.02 & 20 & 2.47 & 1.07 & 2.49 & 1.06\\
\bottomrule
\end{tabular}

}
\caption*{\footnotesize{\textit{Note: Estimates of the average treatment effect using difference-in-means estimator (first column), adjusting for covariates with a linear regression (fifth column), and adjusting for covariates using Augmented Inverse Propensity Weighting - AIPW (column seven); covariates used: users' grade, user type (B2B, B2C, or paid), past utilization, niche type (indicator whether user consumes content that is popular amongst other users or more niche content), past usage of the \emph{Recommended Story} section. Columns two, six, and eight show standard errors. Column three presents p-values. Three first rows describe outcomes in \emph{Recommended Story} section, and three bottom rows overall app utilization.}}}
\end{table}

\begin{figure}[!ht]
        \centering
        \caption{Difference in average \textit{Total Story Engagement}}
        \includegraphics[scale = 0.63]{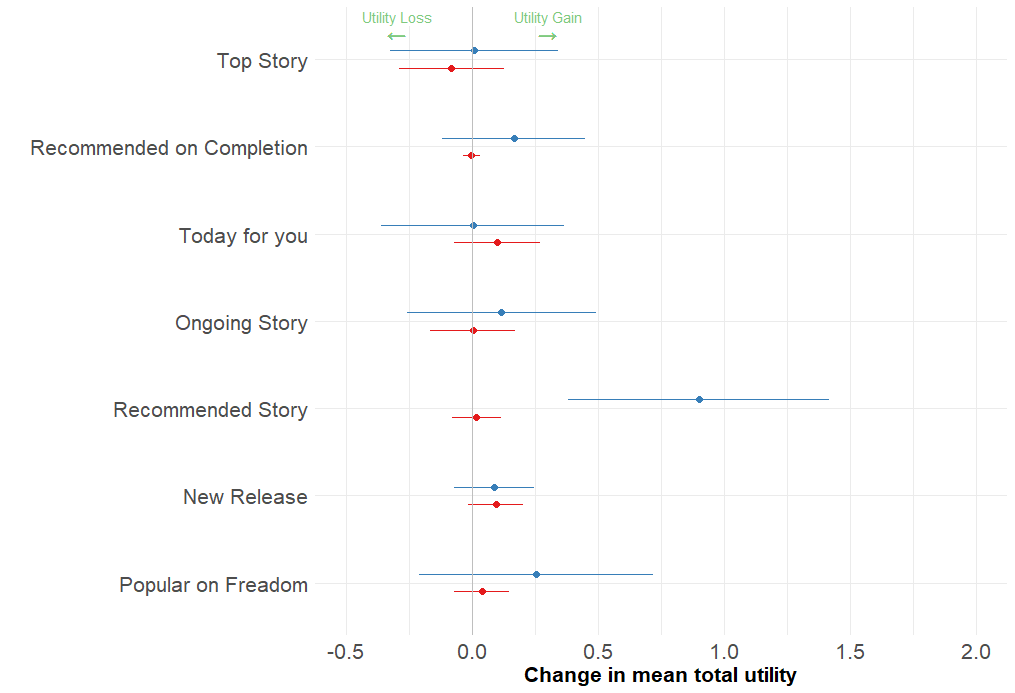}
        \caption*{\footnotesize{\textit{Note: Difference in average \textit{Total Story Engagement} in treatment and control groups for the eight most popular sections. The experimental period is in blue and the pre-experimental period is in red. The pre-experimental period is 7.06.2021 -19.06.2021, and there are approximately twice as many users in the per-experimental period (which was chosen based because it was the closest two-week-long period without other major experiments and alterations in the app).}}}
        \label{fig:ate_all_paths}
    \end{figure}

Additionally, we review differences in \emph{Total Story Engagement} in the most popular sections in the app across treatment and control.  \Cref{fig:ate_all_paths} shows differences in average \emph{Total Story Engagement} in treatment and control groups in other popular sections in the app. The experiment period is marked in blue. From this figure,  we can see that the difference between the two groups is statistically significant only for \emph{Recommended Story} section. We carry out this comparison for the same users in a pre-experiment period. Before the experiment, differences in average Story Engagement across treatment and control are not statistically distinguishable from zero in all of the sections, which is expected since the users were randomly assigned.

\paragraph{Impact on time spent on the app.} As shown in \Cref{tab:ATE}, there is a strong and statistically significant positive effect on the time spent both in \emph{Recommended Story} section as well as across all sections.\footnote{Note, this outcome metric is a sum of the duration of completed stories.  We do not include stories that were started but not completed since we do not observe when users stopped engaging with a specific story. The average number of stories started but not completed in treatment and control is roughly the same, 2.33 in treatment and 2.17 in control; the test for difference in means has a p-value of 0.55.}
    
The increase in total time spent on the app is particularly interesting because it means that students' consumption choices shifted toward using the app and away from other activities. In our context, this result suggests that if the content is interests students, they are willing to go beyond the time prescribed by parents or teachers.
    
Changes in the app quality impact consumption decisions on the intensive and extensive margin. Intensive margin gains are due to users generating higher \emph{Story Engagement} during the time that they spend on the app. This means reallocating the time to more attractive, personalized content in the \emph{Recommended Story} section. The extensive margin impact, on the other hand, implies that users substitute away from other activities and start using the app more. The effect on the extensive margin highlights that the app quality matters to the users, and improving it will result in more time spent with the app.

\paragraph{Mean \emph{Story Engagement}.} We have so far focused on \emph{Total Story Engagement} in the analysis of the treatment effects. \emph{Total Story Engagement} allows us to capture both the increase in the number of user-story interactions and the outcome of those interactions. Now we focus on the average outcomes of the user-story interactions. 

\begin{figure}[!ht]
        \centering
        \caption{Mean \emph{Story Engagement} per rank}
        \includegraphics[scale = 0.55]{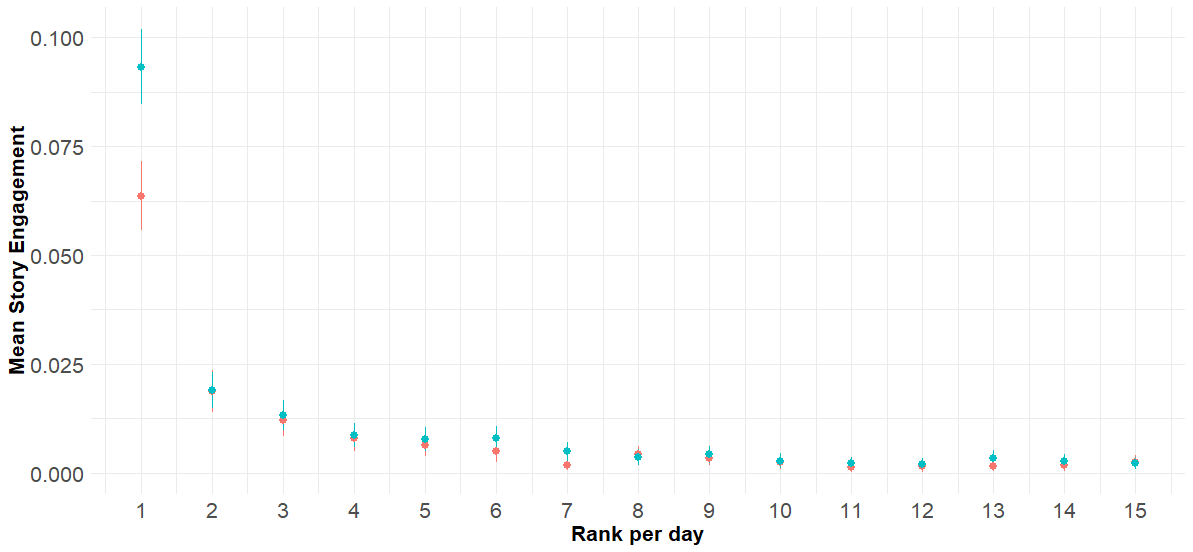}
        \caption*{\footnotesize{\textit{Note: Mean \emph{Story Engagement} per rank per experimental group. In blue we show treatment group and in pink control group. Rank is defined on the basis of the order in which a user interacts with a story on a given day. That is the first story with which a user interacts in a given day gets that rank 1. Whiskers show confidence intervals.}}}
        \label{fig:mean_rank}
    \end{figure}

Defining the outcome as the mean of \emph{Story Engagement} in the experimental group, we estimate the average treatment effect of 0.006 (S.E. 0.013). This effect is statistically not different from zero. This is a puzzling result considering the high average treatment effect on \emph{Total Story Engagement}. Figure \ref{fig:mean_rank} considers the mean \emph{Story Engagement} per rank within a day: the first story that a user interacts with on \emph{Recommended Story} section on a given day has the rank 1, the second rank two, etc. There are 15 ranks in a given day because there are 15 stories displayed in the \emph{Recommended Story} section, so a user cannot interact with more stories in this section on a given day. Stories that were assigned to the user on a day on which a user was active on the platform but did not view them are assigned a \emph{Story Engagement} of 0. The rank we consider here is different from a rank on a slate, where we measure the order of stories displayed to the user; here, we focus on the order of stories that a user has interacted with on a specific day.

Figure \ref{fig:mean_rank}  shows that, first, stories that users interact with first in a day (stories at lower ranks) have substantially lower \emph{Story Engagement} as compared to the stories users interact with later. Second, we find that the mean \emph{Story Engagement} for stories on rank 1 is higher for treatment than for control. The difference between mean \emph{Story Engagement} for treatment and control is statistically insignificant for stories on higher ranks.

\begin{figure}[!ht]
        \centering
        \caption{The number of user-story interactions per experimental group}
        \includegraphics[scale = 0.55]{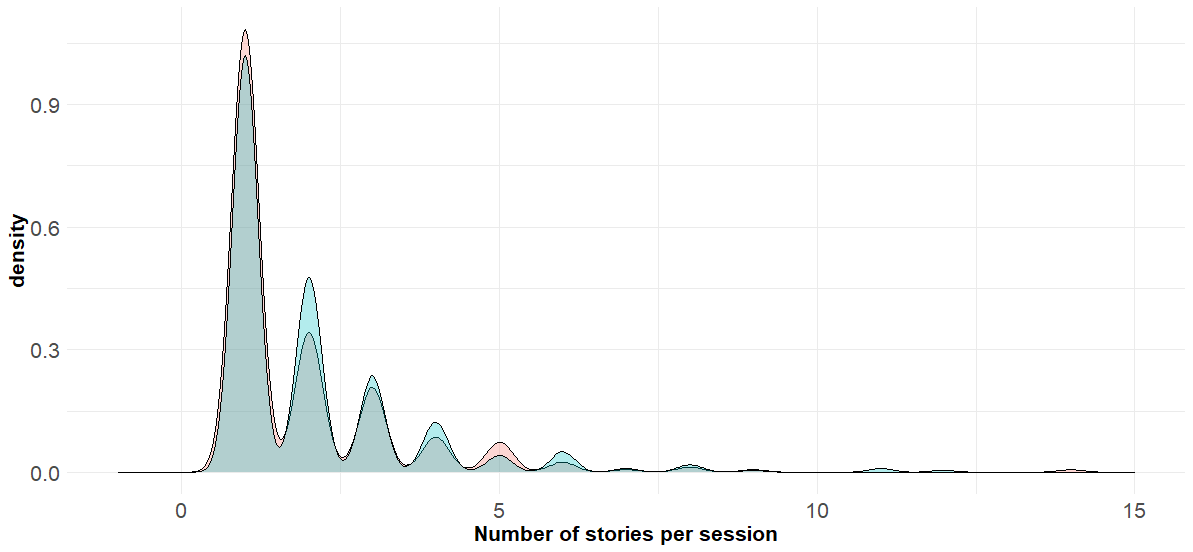}
        \caption*{\footnotesize{\textit{Note: Frequencies of session length per experimental group. Unit of observation is a session. In pink we show control group and in blue treatment group. Session length is measured as the number of stories a user has interacted with.}}}
        \label{fig:no_sessions}
    \end{figure}

Next, we explore the change in the number of user-story interactions per day. We define a new outcome variable: how many story interactions did a user have during a session? Figure \ref{fig:no_sessions} shows a distribution of this outcome. We omit cases in which a user logged in to the app but had zero interactions with stories in the \emph{Recommended Story} tray. We treat individual sessions as a unit of observation. In the control group there were more sessions during which users only interacted with one story. In comparison, in the treatment group there were more sessions in which users interacted with multiple stories.

Putting together evidence from Figure \ref{fig:mean_rank} and Figure \ref{fig:no_sessions}, we find that users in the treatment group had more interactions with stories on higher ranks. Interactions on higher ranks typically lead to lower \emph{Story Engagement}. Hence, the mean \emph{Story Engagement} is not higher in the treatment group despite the high treatment effect on total \emph{Story Engagement}.

\subsection{Heterogeneous treatment effects}\label{HTE_section}

The evidence presented so far relates to the average impact of personalization. In this section, we analyze heterogeneity in treatment across past usage intensity, taste for popular vs. niche content, and the usage of the \emph{Recommended Story} section prior to the experiment. In \Cref{hte_appendix}, we conduct a data-driven analysis of treatment heterogeneity and find moderate treatment heterogeneity. While empirical research using randomized experiments usually documents the average effects of personalization, it is relatively uncommon for studies to analyze heterogeneity in treatment outcomes.\footnote{An exception is the analysis in \citep{SPOTIFYREC}, which most closely resembles our approach.}

We expect that the personalization of content recommendations will benefit heavy and niche-type users the most. Frequent users have a long record of user-story interactions, which allows us to understand their tastes well. Additionally, we expect niche users to have high benefits because whereas in the baseline system, stories are targeted at a typical user, the personalized system caters to their unique tastes.

\paragraph{Definitions of user types.} To determine whether someone is a heavy user, we analyze the pre-experimental app usage. Specifically, for each user, we compute the \emph{Total Story Engagement} and the total number of completed stories prior to the start of the experiment.
Additionally, we construct the following indicator variables: \emph{high engagement user} and \emph{high story completion user}, which take a value of 1 when a user is in the top $50^{th}$ percentile of the distribution of past utilization (past number of completed stories) and 0 otherwise.

\begin{figure}[!ht]
	\centering
			\caption{Histogram of shares of popular stories per user}
			\includegraphics[height=3.5in]{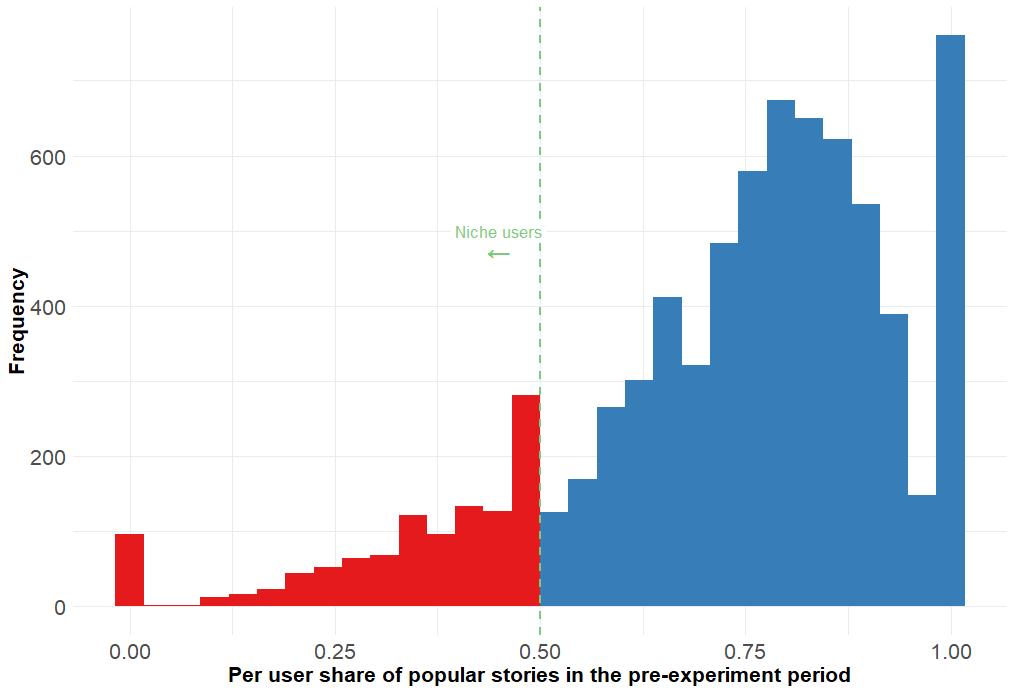}
			\caption*{\footnotesize{\textit{Note: Histogram of per user shares of popular stories in all stories the user had interacted with prior to the experiment. Users are defined as niche if less than 50\% of their interactions were with popular stories. A popular story is a story in the top 25\% of stories ranked by the number of pre-experiment completions. Niche users are in red; non-niche users are in blue.}}}
			\label{fig:hist_popularity}
	\end{figure}

Niche-type users are users that consume content that is generally not very popular. We consider a story to be a popular story if it is one of the top 25\% of stories in terms of pre-experiment completions.\footnote{Top 25\% of stories correspond to 67\% of impressions in the \emph{Recommended Story} during the experiment.}  \Cref{fig:hist_popularity} shows the histogram of shares of popular content consumption per user prior to the experiment. There are some users whose content is largely niche. We consider a user to be a niche type if the share of niche content in her pre-experiment consumption is more than 50\% (in red in \Cref{fig:hist_popularity}). Note that all users received the same recommendations before the experiment, and as such, finding niche stories required searching beyond the top of the recommendation list.

\paragraph{Treatment effects per group.} We start by providing estimates of the average treatment effects per group of interest. We consider total \emph{Story Engagement} in the \emph{Recommended Story} section as the outcome variable of interest and use a difference-in-means estimator. \Cref{tab:HTE_groups} presents the results.

\begin{table}[!htbp] \centering
  \caption{Estimates of average treatment effects by user type}
  \label{tab:HTE_groups}
  \resizebox{0.75\textwidth}{!}{

\begin{tabular}{@{}llccc@{}}
\toprule
\textbf{Statistic} & \textbf{Group}                      & \textbf{Value} & \textbf{S.E.} & \textbf{p-value} \\ \midrule
ATE                & Niche      users                         & 0.334          & 0.089         & <0.01            \\
ATE                & Non-niche       users                    & 0.044          & 0.063         & 0.49             \\
Diff. ATE          & Niche and non-niche    users             & 0.308          & 0.096         & <0.01            \\ \midrule
ATE                & High engagement        users             & 0.299          & 0.094         & <0.01            \\
ATE                & Low engagement     users                 & 0.092          & 0.056         & 0.10             \\
Diff. ATE          & High and low engagement   users          & 0.24           & 0.08          & <0.01            \\ \midrule
ATE                & High completion      users               & 0.249          & 0.094         & <0.01            \\
ATE                & Low completion   users                   & 0.120          & 0.056         & 0.03             \\
Diff. ATE          & High and low completion   users          & 0.153          & 0.081         & 0.03             \\ \midrule
ATE                & High engagement and niche    users      & 0.508          & 0.132         & <0.01            \\
ATE                & High engagement and non-niche    users   & -0.023         & 0.122         & 0.85             \\
Diff. ATE          & High engagement niche and non-niche users & 0.552          & 0.069         & <0.01            \\ \bottomrule
\end{tabular}
}
\caption*{\footnotesize{\textit{Note: Outcome variable is total \emph{Story Engagement} per user. ATE is estimated using a difference-in-means estimator. Diff. ATE refers to difference between the average treatment effects for the two groups.  All groups are defined based on pre-experiment app usage. P-values are for the null-hypotheses that the value is zero.}}}
\end{table}
The gains from personalization are higher for niche users and heavy users compared to non-niche and light users, respectively. The niche average treatment effect is higher and more statistically significant. In the last two rows of \Cref{tab:HTE_groups}, we focus on the distinction between niche and non-niche users in the high-engagement group and find that niche users in this group also have much higher treatment effects. This highlights that the niche users form a distinct category -- i.e., they are not just heavy users who completed all popular stories and need to explore less popular ones.\footnote{One might argue that a user becomes niche after having seen all the popular stories. Note that there are 839 users in the high engagement and niche and 573 in high engagement and non-niche. This indicates that niche users are indeed a distinct category of users.} In \Cref{robust}, we provide further robustness of this result by regressing AIPW scores on past utilization and user type.

\paragraph{Niche-type users see more niche content.} Personalized recommendations benefit niche users because instead of having to search for their favorite stories, the stories are presented to them right at the top of the list of recommended stories. In \Cref{tab:nicher}, we confirm this intuition by comparing the popularity of stories shown to popular and niche types in the treatment and control.

For each story, we compute the share of its impressions in total impressions in an experimental group and rank stories by it (\emph{Rank of impressions}). Additionally, we compute each story's percentile in the distribution of impressions within the experimental group (the total number of stories per experimental group differs).

    \begin{table}
    \begin{center}
    \caption{Type of stories shown by user type and treatment status}\label{tab:nicher}
\begin{tabular}{>{}l|lrrrr}
\toprule
group & variable & mean niche & mean non-niche & std. error & p. value\\
\midrule
  {\textcolor{black}{\textbf{Treatment}}} & Rank of impressions & 379.184 & 343.442 & 17.313 & 0.040\\
  {\textcolor{black}{\textbf{Control}}} & Rank of impressions & 498.730 & 489.391 & 18.357 & 0.611\\
\addlinespace
  {\textcolor{black}{\textbf{Treatment}}} & Percentile of impressions & 0.390 & 0.447 & 0.028 & 0.040\\
  {\textcolor{black}{\textbf{Control}}} & Percentile of impressions & 0.401 & 0.412 & 0.022 & 0.611\\
\bottomrule
\end{tabular}
           			\caption*{\footnotesize{\textit{Note: Type of stories shown to niche and popular-type users across treatment and control. The rank of impressions - stories ranked by the number of impressions during the experiment in the experimental group. Percentile refers to the percentile of the distribution of the share of impressions per story in the total impressions in the experimental group.}}}

	\end{center}
    \end{table}
In the control group, popular and niche-type users see stories of similar popularity, while in the treatment group, niche users are shown more niche stories. This difference in stories shown is statistically significant.

\paragraph{New and old users of \emph{Recommended Story} section.} Another important dimension where heterogeneity may arise is the extent of engagement with the \emph{Recommended Story} section before the experiment. Out of all experimental subjects, only 14\% interacted with at least one story from the \emph{Recommended Story} section in the two weeks prior to the experiment, and 47\% had never interacted with a story in this section.

   \begin{table}
    \begin{center}
    \caption{ATE by past usage of \emph{Recommended Story} section.}\label{tab:new_old}
\begin{tabular}{>{}l|lrrrr}
\toprule
group & variable & ATE & ATE \% baseline & std.error & p.value\\
\midrule
  {\textcolor{black}{\textbf{Past users of RS}}} & Total \emph{Story Engagement} & 0.788 & 54.461 & 0.322 & 0.015\\
  {\textcolor{black}{\textbf{Past users of RS}}} & Total stories & 0.497 & 56.182 & 0.248 & 0.046\\
  {\textcolor{black}{\textbf{Past users of RS}}} & Total time reading & 3.980 & 65.417 & 1.777 & 0.026\\
\addlinespace
  {\textcolor{black}{\textbf{New RS users}}} & Total \emph{Story Engagement} & 0.132 & 87.423 & 0.039 & 0.001\\
  {\textcolor{black}{\textbf{New RS users}}} & Total stories & 0.097 & 143.136 & 0.026 & <0.001\\
  {\textcolor{black}{\textbf{New RS users}}} & Total time reading & 0.692 & 148.707 & 0.189 & <0.001\\
\bottomrule
\end{tabular}
\caption*{\footnotesize{\textit{Note: This figure shows the average treatment effects estimates using Difference-in-Means estimator. Subjects were grouped based on the usage of \emph{Recommended Story} section two weeks prior to the experiment. Three first rows show results for users that viewed at least one story in the section; three bottom rows show users that did not interact with any stories in the \emph{Recommended Story} section in this period. }}}

	\end{center}
    \end{table}

\Cref{tab:new_old} presents estimates of conditional average treatment effects. We consider only outcomes specific to the \emph{Recommended Story} section. Three top rows present results for users that have interacted with at least one story in the \emph{Recommended Story} section in the two weeks prior to the experiment. We find high and statistically significant treatment effects for this group.

The three bottom rows of \Cref{tab:new_old} present the results for users who did not actively use the \emph{Recommended Story} section prior to the experiment. Any user on the app's starting page sees the first few stories in a slate without engaging with it. Thus the effect for users who were not engaging with \emph{Recommended Story} section before has to come from the fact that they like stories displayed at the beginning of the slate. We find that the treatment effects for such users are highly statistically significant and have high point estimates. While the point estimates are small, the percentage change compared to the baseline (usage in the control group) is very high. These results suggest that the introduction of personalized recommendations attracted users to the section who otherwise would not be using it.

\paragraph{Stories that drive the treatment effect.}
Is the increase in total \emph{Story Engagement} driven by a few stories liked by many users or a better assignment of many stories? To answer this question, grouped stories into frequently and rarely shown and compare users' \emph{Story Engagement} in these categories in both the treatment and control groups.

\begin{figure}[H]
    	\caption{Estimates of the conditional expectation of \emph{Story Engagement} per bucket.}
			\centering
			\includegraphics[height=4in]{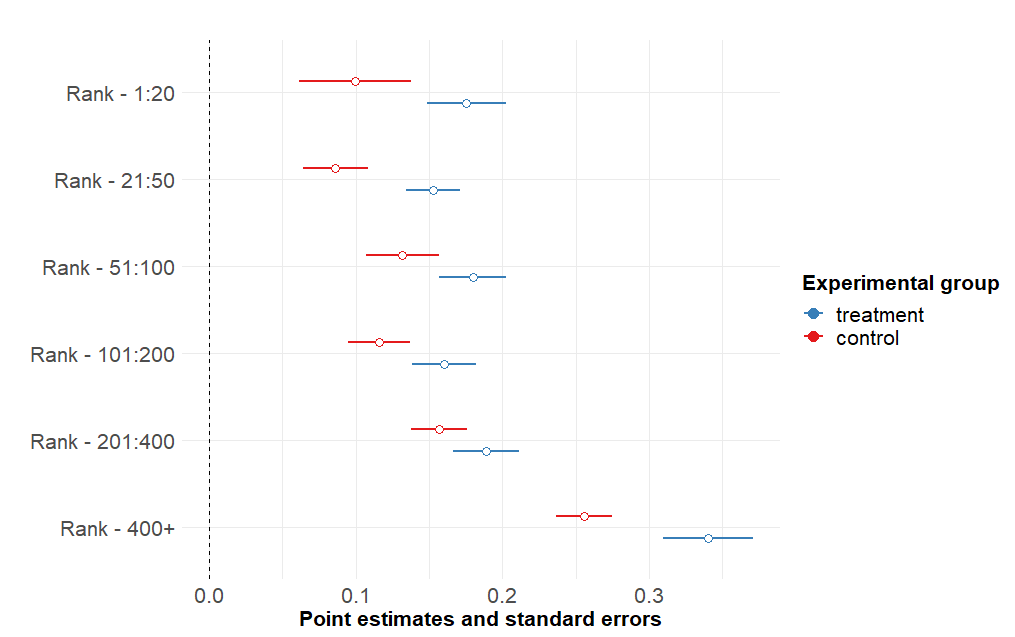}
			\caption*{\footnotesize{\textit{Note: \emph{Story Engagement} estimates adjusted for the difference in grades, user types, and past usage intensity across buckets.}}}
			\label{fig:reg_buckets}
		\end{figure}

\Cref{fig:reg_buckets} shows estimates of the conditional expectation of \emph{Story Engagement} from user-story interactions for stories in different buckets of popularity. We estimate an outcome model using a linear regression framework in which we adjust for users' grades, type, and past utilization. Buckets are constructed according to the rank of the number of story impressions in the experimental group. The total number of impressions in the treatment group is approximately equal in each bucket. Differences across experimental groups in the average \emph{Story Engagement} in a bucket are (apart from personalization) due to the selection of stories into buckets and differences in users that see stories in these buckets. Adjusting for user features allows us to isolate the effect of story selection.

We find that utilities in the treatment group are higher in all buckets. The two first buckets have a high and statistically significant difference between treatment and control.  This suggests that our model picked up stories that many users liked. However, there is also a substantial increase in \emph{Story Engagement} from the least shown, niche stories. This means that there is a component of personalized niche content driving higher \emph{Story Engagement} in the treatment group. In sum, we see that there are two mechanisms in story selection that increase the \emph{Story Engagement} in the treatment group: (i) stories that are shown to many users on average lead to higher \emph{Story Engagement} in the treatment group, and (ii)  personalization of niche, infrequent stories in the treatment group leads on average to higher \emph{Story Engagement} from interactions with these stories.

\section{Conclusion}\label{conclusion}
In this paper, we provide evidence from a randomized controlled trial of the efficacy of personalized recommendations in promoting user engagement on a digital education app. We show that children learning to read in English engage more with content when it is selected based on their preferences. We find an effect of an over 60\% increase in the engagement with the personalized content as compared to the baseline system of content selected by editors. We also find a 15\% boost in overall app usage.

We evaluate the effects of the treatment on different user subgroups in the experiment and find three interesting patterns of heterogeneity. First, we find that heavy users have substantially higher treatment effects. We have more data about such users, which means we know their preferences better and can provide them with higher-quality recommendations. Second, users who ex-ante prefer niche stories are the main beneficiaries of the personalized system. The personalized recommendation system makes it easier for them to discover niche content on the platform. Third, we show that users who, prior to the experiment, were not engaging with stories in the section of the app in which we introduced personalized recommendations reaped high benefits from the new system.

This paper contributes to the recommendation systems literature by bringing evidence from the educational sector and a setting with limited data (as compared to big-tech environments where such systems are typically deployed). We carefully discuss the recommendation system design process hoping to allow practitioners to develop and deploy similar recommendation systems in other contexts.

The main limitation of this paper is that we focus on students that are heavy app users (those who interacted with at least sixty stories) and on stories that have already been shown to many users. This is a limitation of any system based on the matrix factorization model, as the model's performance improves with the number of past user-content interactions. Furthermore, the approach is not applicable to new users and new stories. Developing and implementing recommendations for new users and new items are valuable extensions of this work.

A further limitation is that the proposed approach optimizes for user engagement rather than for learning. The recommendation system assigns stories the user is most likely to complete, but these might not necessarily be the stories that will maximize learning. While optimizing the story selection for learning would be a preferable approach, we focused on engagement because of difficulties in accurately measuring learning outcomes and slower feedback loops. This relates to the literature on surrogate \citep{surrogates_eckles}, where a surrogate metric that closely tracks the target metric is optimized instead due to the target metric being infeasible to access. Bridging the gap between optimizing for short-term outcomes vs. long-term learning, for example, by using surrogates, is a promising next step on this research agenda.

\newpage

\bibliographystyle{apalike}
\bibliography{references}

\newpage{}
\setcounter{page}{1}
\gdef\thepage{A\arabic{page}}
\appendix
\section*{Appendix}

\section{Covariate balance check}\label{cobal}

Table \ref{tab:baltab} presents comparison of means of user characteristics across treatment and control. We find that difference between treatment and control are small and statistically insignificant.

\begin{table}[!htbp] \centering 
  \caption{Balance of covariates across treatment and control} 
  \label{tab:baltab} 
\resizebox{0.75\textwidth}{!}{%
\begin{tabular}{>{}l|rrrrr}
\toprule
covariate & mean treatment & sd treatment & mean control & sd control & p value\\
\midrule
  {\textcolor{black}{\textbf{past Story Engagement}}} & 101.23 & 139.38 & 94.67 & 114.35 & 0.17\\
  {\textcolor{black}{\textbf{past stories}}} & 57.14 & 89.03 & 52.75 & 77.22 & 0.16\\
  {\textcolor{black}{\textbf{max streak}}} & 18.58 & 85.00 & 17.79 & 84.00 & 0.80\\
  {\textcolor{black}{\textbf{share b2b}}} & 0.31 & 0.46 & 0.29 & 0.46 & 0.46\\
  {\textcolor{black}{\textbf{share b2c}}} & 0.41 & 0.49 & 0.41 & 0.49 & 0.79\\

  {\textcolor{black}{\textbf{share paid}}} & 0.25 & 0.43 & 0.26 & 0.44 & 0.66\\
  {\textcolor{black}{\textbf{share grade 2}}} & 0.24 & 0.43 & 0.24 & 0.43 & 0.82\\
  {\textcolor{black}{\textbf{share grade 3}}} & 0.22 & 0.42 & 0.22 & 0.41 & 0.69\\
\bottomrule
\end{tabular}
}
\caption*{\footnotesize{\textit{Note: Means of users' characteristics in treatment and control. Last column p-value from a t.test for difference in means. Category paid includes users from a paid fLive program and regular paying users; category b2b includes regular b2b customers and club 1br users, a B2B promotion.}}}
\end{table}

\section{Robustness check of the average treatment effect estimates}\label{ap_out}
In table \ref{tab:ATE_rob} we present estimates of the average treatment effect based on data which is trimmed at the 95\% percentile of daily stories completed, i.e., we remove users that are in top 5\% of users with highest daily number of completed stories across all paths of the app.

We find very similar estimates of the ATE for outcomes across all paths in the app. The path specific estimates are smaller, but still high and statistically significant. The confidence intervals include the point estimates from the baseline specification.

\begin{table}[!htbp] \centering 
  \caption{Estimates of average treatment effects for all outcome variables} 
  \label{tab:ATE_rob} 
\resizebox{0.8\textwidth}{!}{%
\begin{tabular}{>{}l|rrrr}
\toprule
variable & ATE & std.error & p.value & ATE percentage\\
\midrule
  {\textcolor{black}{\textbf{Total Story Engagement RS}}} & 0.14 & 0.04 & <0.001 & 58\\
  {\textcolor{black}{\textbf{Total stories completed RS}}} & 0.09 & 0.03 & <0.001 & 68\\
  {\textcolor{black}{\textbf{Total reading time RS}}} & 0.67 & 0.21 & <0.001 & 77\\
\addlinespace
  {\textcolor{black}{\textbf{Total Story Engagement all paths}}} & 0.50 & 0.23 & 0.03 & 15\\
  {\textcolor{black}{\textbf{Total stories completed all paths}}} & 0.32 & 0.13 & 0.01 & 21\\
  {\textcolor{black}{\textbf{Total reading time all paths}}} & 2.07 & 0.88 & 0.02 & 20\\
\bottomrule
\end{tabular}
}
\caption*{\footnotesize{\textit{Note: Estimates of the average treatment effect using difference-in-means estimator. Three first rows describe outcomes in \emph{Recommended Story} path, three bottom rows overall app utilization. Last columns shows the ATE estimate as a percent share of the baseline.}}}
\end{table}

\section{Heterogeneous treatment effects}\label{robust}
\begin{figure}%
    \caption{AIPW scores across past utilization. AIPW scores for users with past \emph{Story Engagement} higher than the percentile.}%
    \centering
    \subfloat[\centering Past story completions. AIPW scores for users with the past number of completions higher than the percentile.]{{\includegraphics[width=15cm]{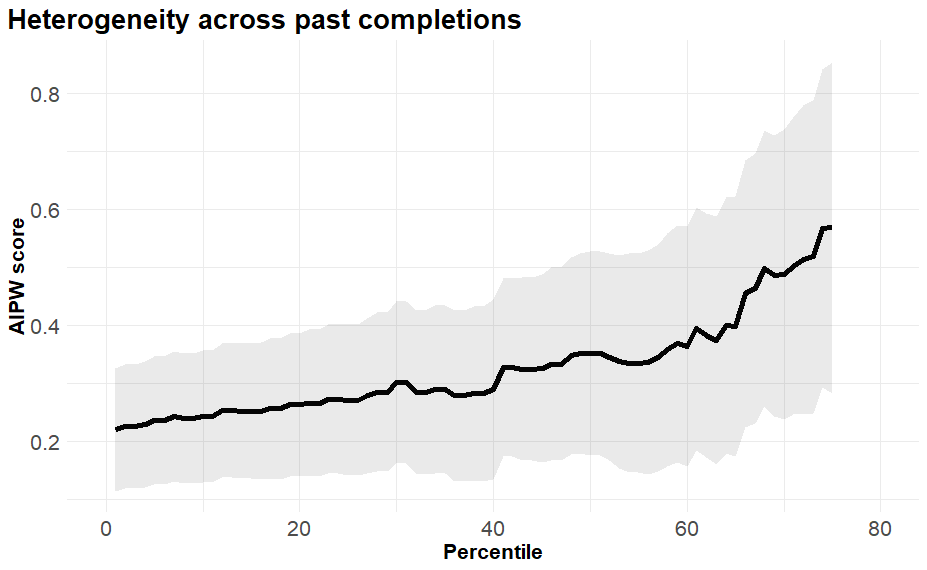} }}%
    \qquad
    \subfloat[\centering Past \emph{Story Engagement}.  AIPW scores for users with past \emph{Story Engagement} higher than the percentile. ]{{\includegraphics[width=15cm]{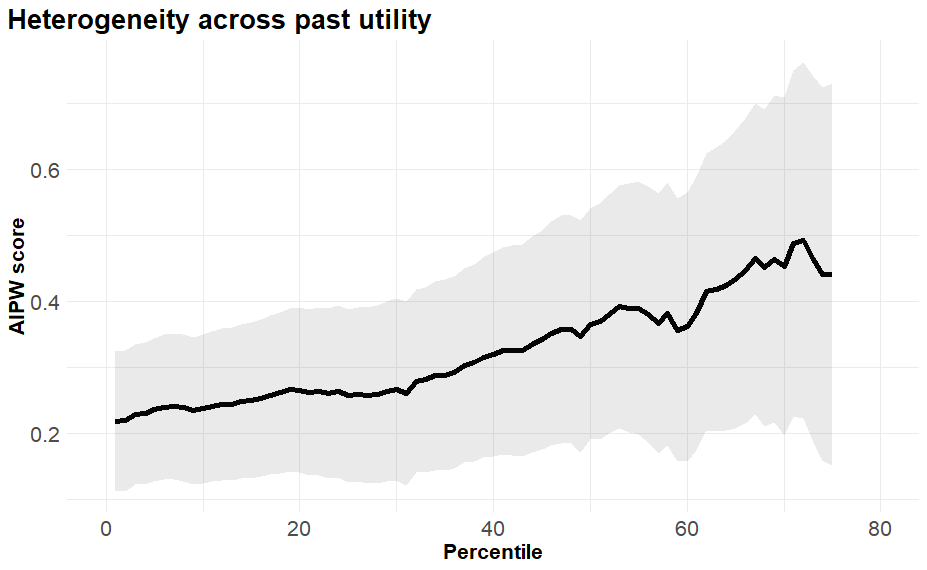} }}%
    \label{fig:aipw_utilization}%
\end{figure}

In \Cref{fig:aipw_utilization}, we show how AIPW scores change across users depending on their past utilization.\footnote{To estimate AIPW scores we use the \emph{grf} package (see \cite{athey2019generalized}). This methodology allows us to adjust flexibly for individual characteristics and estimate conditional average treatment effects. We consider users' school grade, type (B2B, B2C, paid), max streak (maximal number of consecutive days in which users completed at least one story), past utilization (the total number of completed stories prior to the experiment, and total \emph{Story Engagement} prior to the experiment), and whether a user is a niche type. To determine the variables based on past consumption, we consider a period of app usage between July 2020 and the start of the experiment.} Panel A shows how AIPW scores change depending on the percentile of the pre-experiment number of story completions and panel B on users' past \emph{Story Engagement}. Upward trends are apparent in both figures. The differences are, however, moderate.

To provide further robustness into the finding that heavy and niche users benefit from personalized recommendations, we present results of regressions of AIPW scores based on total Story Engagement on users' past utilization (see \cite{athey2019estimating} for methodology). See \cref{tab:aipw_reg} for summary of results. 

\begin{table}[!htbp] \centering 
  \caption{Results of a regression of user types on AIPW scores.} 
  \label{tab:aipw_reg} 
    \resizebox{\textwidth}{!}{
\begin{tabular}{@{\extracolsep{5pt}}lccccccc} 
\\[-1.8ex]\hline 
\hline \\[-1.8ex] 
 & \multicolumn{7}{c}{\textit{Dependent variable:}} \\ 
\cline{2-8} 
\\[-1.8ex] & \multicolumn{7}{c}{Total Story Engagement (aipw.scores)} \\ 
\\[-1.8ex] & (1) & (2) & (3) & (4) & (5) & (6) & (7)\\ 
\hline \\[-1.8ex] 
 past Story Engagement & 0.001$^{***}$ (0.0003) &  &  &  &  &  & 0.001 (0.0004) \\ 
  stories completed &  & 0.002$^{***}$ (0.001) &  &  &  & 0.001$^{**}$ (0.001) &  \\ 
  heavy user Story Engagement &  &  & 0.366$^{***}$ (0.075) &  &  &  &  \\ 
  heavy user completions &  &  &  & 0.352$^{***}$ (0.076) &  &  &  \\ 
  niche type &  &  &  &  & 0.342$^{***}$ (0.074) & 0.231$^{***}$ (0.088) & 0.256$^{***}$ (0.091) \\ 
 \hline \\[-1.8ex] 
Observations & 2,661 & 2,661 & 2,661 & 2,661 & 2,661 & 2,661 & 2,661 \\ 
R$^{2}$ & 0.006 & 0.007 & 0.009 & 0.008 & 0.008 & 0.010 & 0.009 \\ 
\hline 
\hline \\[-1.8ex] 
\textit{Note:}  & \multicolumn{7}{r}{$^{*}$p$<$0.1; $^{**}$p$<$0.05; $^{***}$p$<$0.01} \\ 
\end{tabular} 
}
\caption*{\footnotesize{\textit{Note: Outcome variable is AIPW score of total Story Engagement per user. OLS estimator. All covariates defined based on the pre-experiment app usage. $^{*}$p$<$0.1; $^{**}$p$<$0.05; $^{***}$p$<$0.01}}}
\end{table} 

Columns one to four of \cref{tab:aipw_reg} show that users with high past utilization have higher treatment effects. Column five shows higher treatment effect for niche type users. Finally, columns six and seven control both for heavy utilization and niche type; niche type remains to have a high and statistically significant treatment effect.

\section{Data-driven treatment effects heterogeneity}\label{hte_appendix}

We use the estimated causal forest to divide our users into tertiles according to their estimates CATE prediction (see \cite{chernozhukov2018generic} for details of this approach). To avoid using model that was fitted using observations for which we make predictions, we use honest sample splitting with 10 folds. 

\begin{figure}[H]
			\centering
			\caption{Average CATE within each ranking (as defined by predicted CATE). Predictions with OLS in blue and AIPW scores in red.}
			\includegraphics[height=3.5in]{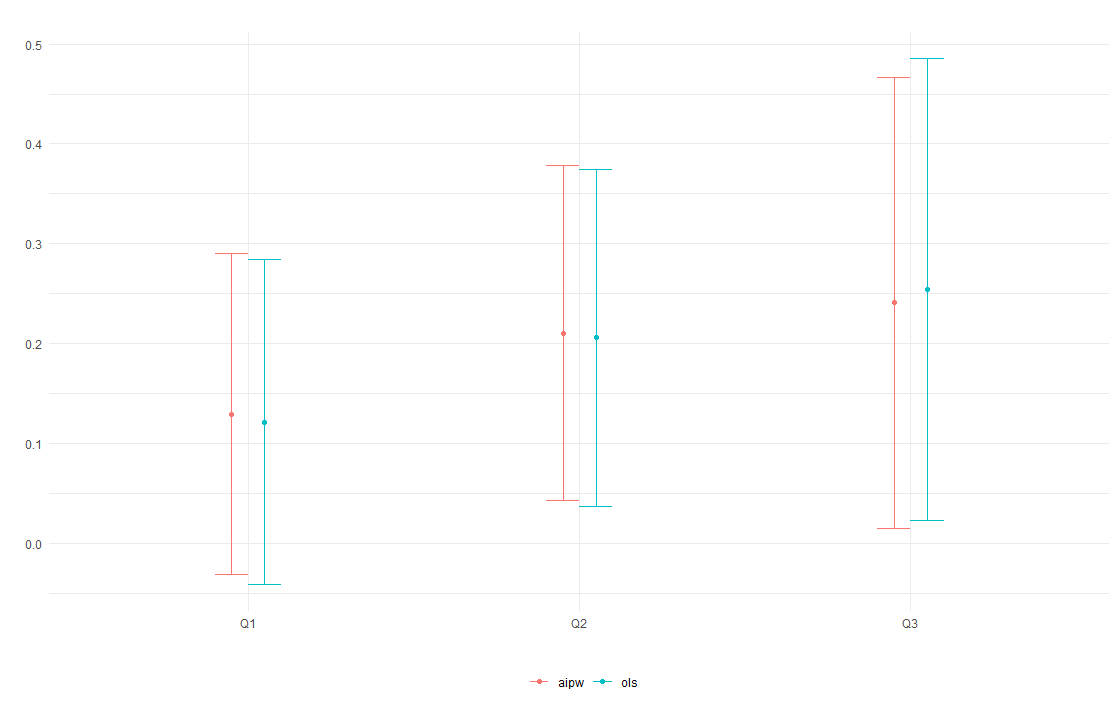}
			\label{fig:CATE_4}
	\end{figure}

\Cref{fig:CATE_4} shows the the predicted CATES in the four groups. First of all, the treatment effects are quite similar for the four groups. The fourth quartile appears to have higher treatment effects, but the differences are small.

\begin{figure}[H]
			\centering
			\caption{Average covariate values within group (based on CATE estimate ranking).}
			\includegraphics[height=5in]{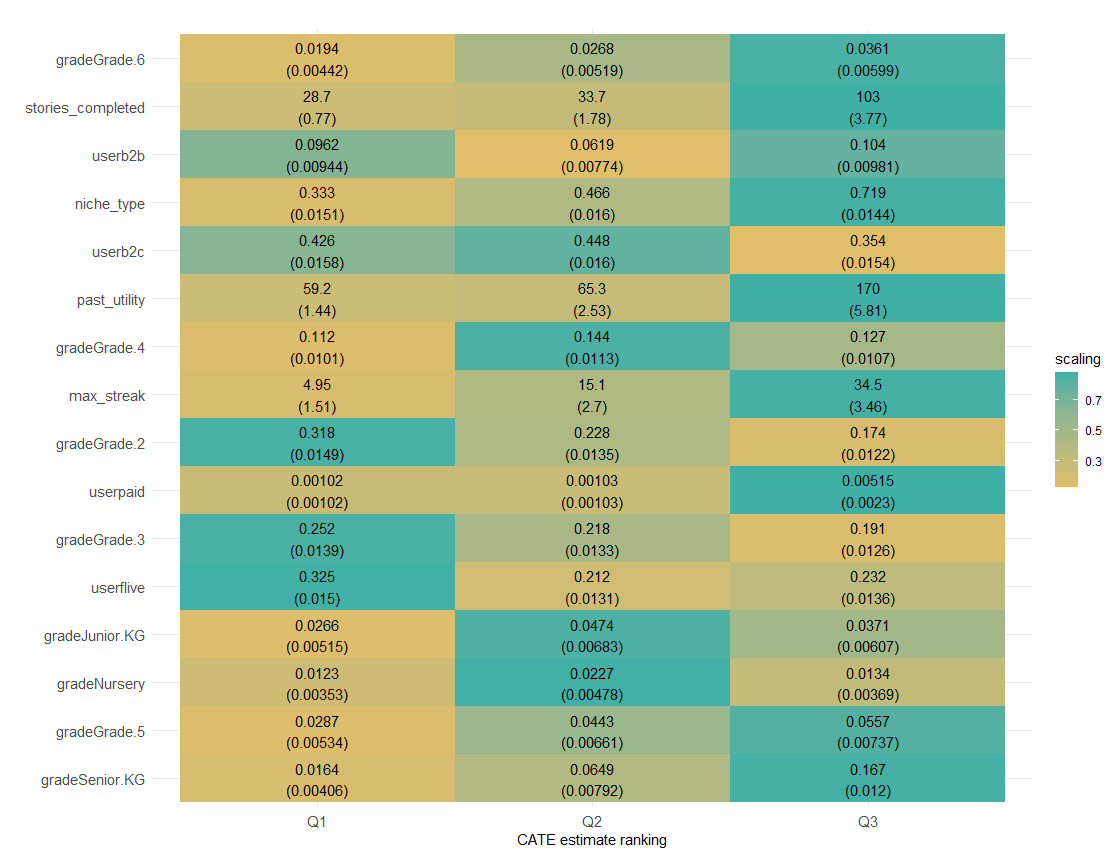}
			\label{fig:average_CATEs}
	\end{figure}

Finally, we can also compare average characteristics for individuals in the four quartiles. We present such a comparison in \Cref{fig:average_CATEs}.

Heavy users (high maximal streak and freq-user indicator) appear more frequently in the highest quartile. We also see more niche users in the fourth group. We look in detail into these groups in the next subsections.

\section{Model calibration with experimental data}\label{calibration}
\begin{figure}
    \centering
    \caption{Histogram of predicted Story Engagement}
    \includegraphics[scale = 0.6]{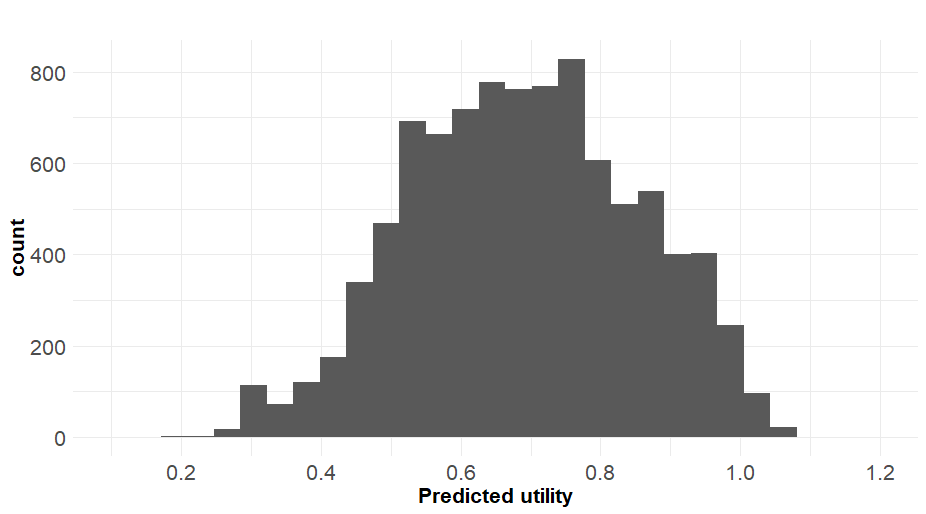}
    \label{hist}
\end{figure}

The main component of the recommendation system is the collaborative filtering model that predicts user Story Engagement from user-story interactions. In our analysis, high treatment effects suggest that the model has successfully identified user preferences and selected stories that users liked.'
In this section, we further evaluate the calibration of the model by correlating the models predicted user utilities with observed utilities from the experiment. Figure \ref{hist} shows the histogram of the predicted utilities; we can notice that they vary from very high values of around 1 to lower values of $0.2$. We don't see values lower than 0.2 because in the experiment we considered only stories ranked at the top of the ranking of predicted utilities.

To evaluate the model calibration, we regress the predicted Story Engagement on the observed Story Engagement from the experiment. Regression results are in Table \ref{tab:ev_pred_all}.

\begin{table}[!htbp] \centering 
                \caption{Correlation between Story Engagement predicted by the collaborative filtering model and observed in the experiment. $^{***}: p < 0.01$} 
\begin{tabular}{@{\extracolsep{5pt}}lccc} 
\\[-1.8ex]\hline 
\hline \\[-1.8ex] 
 & \multicolumn{3}{c}{\textit{Dependent variable:} Story Engagement} \\ 
\cline{2-4} 
 & All users & Frequent users & Infrequent users \\ 
\hline \\[-1.8ex] 
pred. Story Engagement & 0.44$^{***}$ (0.01) & 0.46$^{***}$ (0.01) & 0.43$^{***}$ (0.01) \\ 
 \hline \\[-1.8ex] 
Observations & 9,344 & 2,268 & 7,076 \\ 
R$^{2}$ & 0.40 & 0.38 & 0.41 \\ 
\hline 
\hline \\[-1.8ex] 
\end{tabular}
         \label{tab:ev_pred_all} 
         
        \end{table}
 \Cref{tab:ev_pred_all} shows that the model predictions are strongly correlated with observed utilities. The model is better calibrated for frequent users, for whom we have longer consumption histories (albeit the difference is small). We also break down the analysis by the experimental groups, which is shown in \Cref{tab:ev_pred_groups}.
    
\begin{table}[!htp] \centering 
 \caption{Results from linear regressions of actual Story Engagement on predicted Story Engagement. Column (1) treatment group, column (2) control group.} 
\begin{tabular}{@{\extracolsep{5pt}}lcc} 
\\[-1.8ex]\hline 
\hline \\[-1.8ex] 
 & \multicolumn{2}{c}{\textit{Dependent variable:} Story Engagement} \\ 
\cline{2-3} 
 & treatment & control \\
\hline \\[-1.8ex] 
 pred. Story Engagement & 0.33$^{***}$ (0.01) & 0.60$^{***}$ (0.01) \\ 
 \hline \\[-1.8ex] 
Observations & 4,695 & 1,773 \\ 
R$^{2}$ & 0.32 & 0.51 \\ 
\hline 
\hline \\[-1.8ex] 
\end{tabular} 
         \label{tab:ev_pred_groups} 
\end{table} 

In \Cref{tab:ev_pred_groups} we can notice that the predictions from the model correlate strongly with observed Story Engagement in both treatment and control groups. The model is much better calibrated in the control group, this is not surprising because the model is trained on similar data

\end{document}